\documentclass[fleqn,usenatbib]{mnras}
\usepackage{newtxtext,newtxmath}
\usepackage[T1]{fontenc}
\usepackage{ae,aecompl}
\usepackage{adjustbox}
\usepackage{graphicx}
\usepackage{epsfig}

\usepackage{amssymb}	
\usepackage{amsmath}	
\usepackage{natbib}
\usepackage{threeparttable}
\usepackage{booktabs}
\usepackage{xcolor}
\usepackage{multirow}
\usepackage{soul}
\usepackage{subfigure}
\usepackage{multirow}
\usepackage{setspace}
\usepackage{longtable}
\usepackage{multicol}
\usepackage{epstopdf}
\usepackage{times}
\usepackage{float}

\newcommand{\swift}{\textit{Swift}}

\newcommand{\source}{Swift~J1753.5-0127}

\def \nar {NewAR}
\def \aj {AJ}
\def \mnras {MNRAS}
\def \apj {ApJ}
\def \apjs {ApJS}
\def \apjl {ApJL}
\def \aap {A\&A}

\def \araa {ARAA}

\def \pasp {PASP}
\def \pasj {PASJ}

\def \aapr {A\&A Rev.}

\title[The origin of UV/optical emission in \source]{The origin of UV/optical emission in the black hole low-mass X-ray binary \source}

\author[\rm Pengcheng Yang et al.]{
Pengcheng Yang$^{1,2,3,4}$\thanks{E-mail: yangpengcheng@ynao.ac.cn}, Guobao Zhang$^{1,2,3,4}$\thanks{E-mail: zhangguobao@ynao.ac.cn}, David M. Russell$^{5,6}$, Joseph D. Gelfand$^{5,6,7}$, \newauthor{Mariano M\'endez $^{8}$, Jiancheng Wang$^{1,2,3,4}$ and Ming Lyu$^{9,10}$}
\\
$^{1}$Yunnan Observatories, Chinese Academy of Sciences, Kunming 650216, People's Republic of China \\
$^{2}$Key Laboratory for the Structure and Evolution Celestial Objects, Chinese Academy of Sciences, Kunming 650216, People's Republic of China\\
$^{3}$Center for Astronomical Mega-Science, Chinese Academy of Sciences, Beijing 100012, People's Republic of China \\
$^{4}$University of Chinese Academy of Sciences, Beijing 100049, People's Republic of China  \\
$^{5}$New York University Abu Dhabi, P.O. Box 129188, Abu Dhabi, UAE\\
$^{6}$Center for Astro, Particle and Planetary Physics, New York University Abu Dhabi, P.O. Box 129188, Abu Dhabi, UAE\\
$^{7}$Center for Cosmology and Particle Physics, New York University, 726 Broadway, Room 958, New York, NY 10003, USA\\
$^{8}$Kapteyn Astronomical Institute, University of Groningen, P.O. Box 800, NL-9700 AV Groningen, The Netherlands     \\
$^{9}$Department of Physics, Xiangtan University, Xiangtan, Hunan 411105, People's Republic of China    \\
$^{10}$Key Laboratory of Stars and Interstellar Medium, Xiangtan University, Xiangtan, Hunan 411105, People's Republic of China
}

\date{Accepted 2022 April 16. Received 2022 April 12; in original form 2021 November 2}

\pubyear{2022}

\begin{document}
\label{firstpage}
\pagerange{\pageref{firstpage}--\pageref{lastpage}}
\maketitle

\begin{abstract}
The emission from the accreting black holes (BHs) in low-mass X-ray binaries (LMXBs) covers a broad energy band from radio to X-rays. Studying the correlations between emission in different energy bands during outbursts can provide valuable information about the accretion process. We analyse the simultaneous optical, ultraviolet (UV) and X-ray data of the BH-LMXB \source\ during its $\sim12-$year long outburst with the \textit{Neil Gehrels Swift Observatory}. We find that the UV/optical and X-ray emission are strongly correlated during the hard states of the outburst. We fit the relation with a power-law function $F_{\rm{UV/optical}} \propto F_{X}^{\beta}$ and find that the power-law index $\beta$ increases from $\sim0.24$ to  $\sim0.33$ as the UV/optical wavelength decreases from $\sim$ 5400 \AA\ (V) to $\sim$ 2030 \AA\ (UVW2). We explore the possible reasons for this and suggest that in \source\ the UV/optical emission is dominated by a viscously heated accretion disc at large radii. We find that the data that deviate from the correlation correspond to the low-intensity peaks appeared in the X-ray band during the outburst, and suggest that these deviations are driven by the emission from the inner part of the accretion disc.

\end{abstract}



\section{Introduction}
\label{sec:introduction}

Low-mass X-ray binaries (LMXBs) are binary systems in which a compact object, a black hole (BH) or a neutron star (NS), accretes mater from a low-mass ($< 1 M_\odot$ ) non-collapsed companion star via an accretion disc. Most LMXBs are X-ray transient systems, spending most of their lifetime in quiescence at a low X-ray luminosity. After a quiescent state, occasionally, these X-ray transient systems experience outbursts. During an outburst, the X-ray and optical flux increase by several orders of magnitude and even approach the Eddington limit \citep{1997ApJ...491..312C}. For a typical LMXB outburst, the light curve shows a fast rise and an exponential decay \citep[FRED,][]{1997ApJ...491..312C}. 

The physical mechanism behind LXMB outbursts could be explained by a thermal-viscous instability in the accretion disc. As described in the disc instability model \citep[DIM, for a review, see][]{2001NewAR..45..449L}, the mater from the companion star accumulate in the outer accretion disc during quiescence; eventually, the surface density reaches a critical point, and the material can be heated and ionized. In a heated, ionized disc, as the viscosity of the disc increases, the angular momentum is more easily transferred outward, resulting in the mater to rapidly flowing onto the compact object and triggering the outburst\citep{2001NewAR..45..449L,2001A&A...373..251D,2020AdSpR..66.1004H}.

During these outbursts, most BH-LMXBs show similar q-shaped tracks in the hardness-intensity diagram \citep[HID,][]{2001ApJS..132..377H, 2004MNRAS.355.1105F}. BH-LMXBs go through a number of different spectral states (e.g. hard, soft, intermediate, quiescent), which are classified by the X-ray spectral and timing properties. The hard state (HS) is usually observed at the beginning and the end of a typical outburst, where the X-ray spectrum is dominated by the non-thermal emission in the form of a power-law with a photon index of $1.4 < \Gamma < 2.1$ and a high-energy cut-off at $\sim100$ keV \citep{2006ARA&A..44...49R}. The physical origin of this non-thermal component is debated, which is commonly believed that it is due to inverse Compton scattering of soft photons in an optically thin corona consisting of hot electrons\citep{1980A&A....86..121S, 1994ApJ...434..570T, 2007A&ARv..15....1D, 2017MNRAS.466..194B}. In the soft state (SS), the spectrum is dominated by the thermal emission from an geometrically thin and optically thick accretion disc \citep{1973A&A....24..337S}.
When the source moves from (to) the HS to (from) the SS, the transitional states are classified as the hard intermediate state (HIMS) and the soft intermediate state (SIMS); the spectral hardness of the source is in between the values observed in the HS and SS. In general, during the outburst, the sequence of states from quiescence to quiescence is the following: HS - HIMS - SIMS - SS - SIMS - HIMS - HS (See classifications and characteristics of spectral states in \citet{2006ARA&A..44...49R, 2016ASSL..440...61B} for more details). However, there are sources that during some outbursts remained in the hard states (HS/HIMS) and never evolved into the softer states (SIMS/SS), such as the 2000 outburst of XTE J1118+480 \citep{2001ApJ...563..246W}, the 2010 outburst of MAXI J1659-152 \citep{2015ApJ...803...59D}, the 2011 outburst of MAXI J1836-194 \citep{2016ApJ...819..107J}, the 2011 and 2017 outburst of Swift J1357.2-0933 \citep{2013MNRAS.428.3083A, 2019MNRAS.485.3064B}. These types of "failed-transition" outbursts are considered to be common events in BH-LMXBs \citep{2016ApJS..222...15T, 2021MNRAS.507.5507A}.

\begin{figure*}
\begin{center}
\includegraphics[scale=.54]{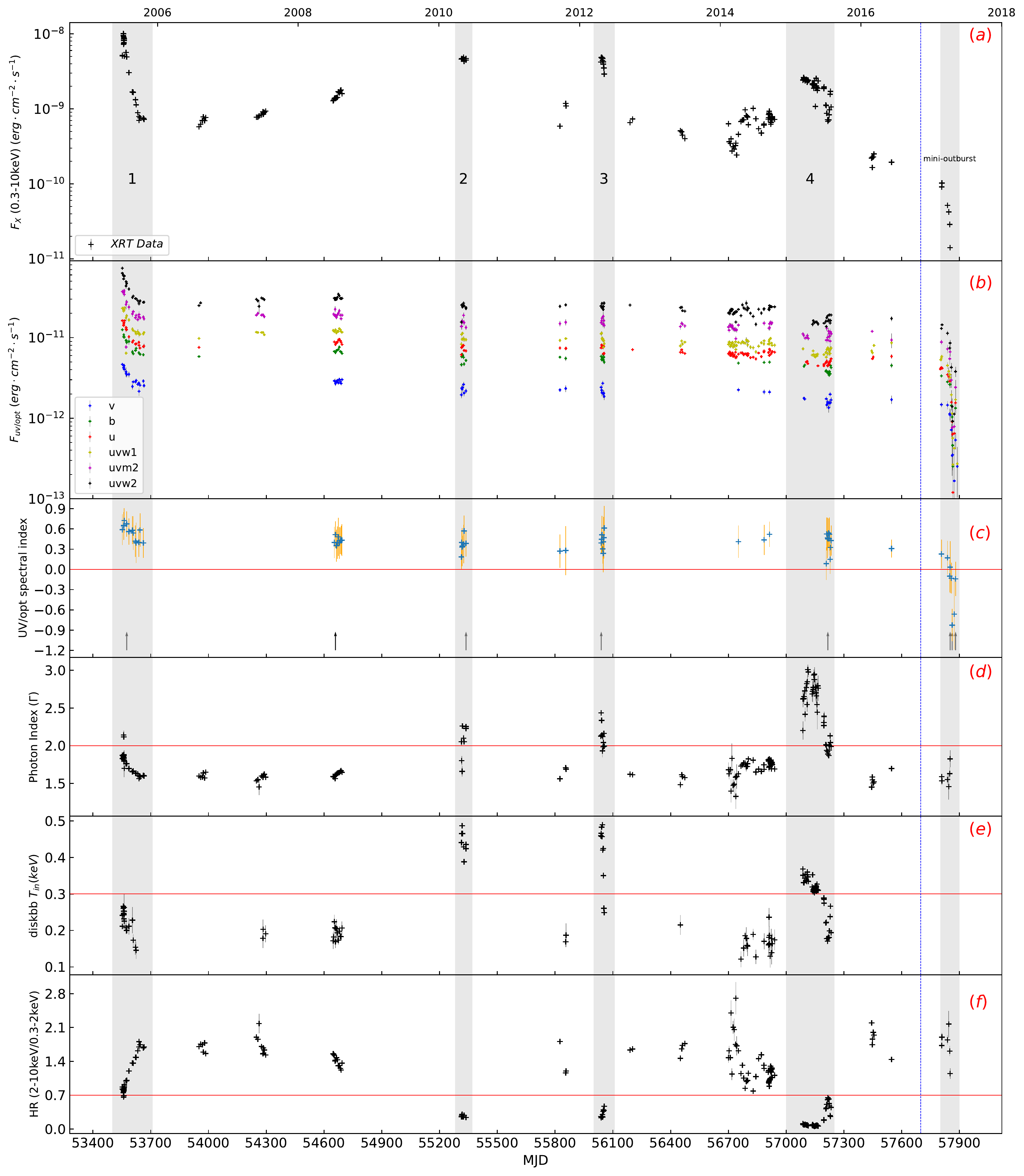}
\caption{Outburst observation of \source with \swift. Panel (a): unabsorbed X-ray (0.3$-$10 keV) light curve. Panel (b): dereddened UV/optical light curves in six UVOT filters. Panel (c): the spectral index of the UV/optical SED; black arrows mark where we select the representative SED. Panels (d), (e) and (f): photon index ($\Gamma$), temperature at the inner disc radius ($T_{in}$) and hardness ratio (HR) as a function of time. Each point represents one observation. The first gray region represents the early phase of the main outburst. The gray regions 2, 3, 4 are low-intensity peaks, in which the region 2 is the failed transition from \citet{2013MNRAS.429.1244S}, the region 3 is the second "dip" from \citet{2013MNRAS.433..740S}, and the region 4 is the low-luminosity soft state from \citet{2016MNRAS.458.1636S}. The last gray area denotes the mini-outburst. The blue dashed line indicates when the source went into quiescence \citep{2016ATel.9739....1A,2016ATel.9765....1P,2016ATel.9708....1R} before the first mini-outburst.
These horizontal red lines are the reference lines defined in the main text (see Sec.\ref{subsec:lc}, Sec.\ref{subsec:sed} and Sec.\ref{subsec:x_uv/opt correlation}).}
\label{fig:para_time}
\end{center}
\end{figure*}

The emissions from the BH-LMXBs cover a broad energy band from radio to X-ray. Therefore, multi-wavelength observations can be used to study the source of radiation at different wavelengths. The radio emission from LMXBs is believed to be the synchrotron process in jets  \citep{1995xrbi.nasa..308H}. However, though the X-ray and radio emission processes in BH-LMXBs are relatively well-understood, the origin of the emission in the ultraviolet (UV) and optical bands is not clearly known yet. 

Linking the fluxes produced at different wavelengths could provide valuable information about the accretion mechanism. A useful method has been applied to analyse the origin of UV/optical radiation by studying the power-law correlation, $L_{\rm{UV/optical}} \propto L_{X}^{\beta}$, between the UV/optical and the X-ray emission. The power-law index $\beta$ can provide some constraints on the possible emission mechanisms. Several possible models to explain the UV/optical emission of BH-LMXBs in the hard state have been proposed according to the value of $\beta$ :

\begin{enumerate}
\item The UV/optical emission of LMXBs is the result of the reprocessing of X-ray in the outer accretion disc \citep{1994A&A...290..133V}. This model is based on the simple geometric assumption that the accretion disc is geometrically thin and optically thick. The outer region of the accretion disc intercepts and reprocesses a portion of the X-ray flux from the central source. The temperature ($T$) of the outer accretion disc is dominated by X-ray irradiation, and the irradiated outer disc contributes to the UV/optical emission. In this model there is an empirical relation between the V-band and the X-ray luminosity: $L_{\rm{V}} \propto L_{\rm{X}}^{\beta}$, where $\beta$ depends on the surface brightness of the accretion disc given by $S_{\nu} \propto T^{\alpha}$, $\beta = \alpha / 4$, and in the V-band $\beta \sim 0.5$ ($\alpha \sim 2.0$) \citep{1994A&A...290..133V}. \citet{2015MNRAS.453.3461S} also pointed out that $\beta$ changes with the frequency bands as: in the UV band, $\beta \sim 0.9$ ($\alpha \sim 3.7$), in the K-band, $\beta \sim 0.3$ ($\alpha \sim 1.2$).

\item The UV/optical emission originates from the synchrotron radiation in the jets \citep{2006MNRAS.371.1334R}. For BH-LMXBs, the jet usually exists in the hard state and is quenched in the soft state \citep{2004MNRAS.355.1105F}. The total jet power scales with the radio luminosity and mass accretion rate ($\dot{m}$), following the relation $L_{\rm{radio}} \propto L_{\rm{jet}}^{1.4}$\citep{1979ApJ...232...34B,1996A&A...308..321F,2001A&A...372L..25M,2003MNRAS.343L..59H}, $L_{\rm{jet}} \propto \dot{m}$ \citep{1996A&A...308..321F,2006MNRAS.369.1451K}. For radiatively inefficient accretion, such as BH-LMXBs in the hard state, the X-ray luminosity scales as $L_{\rm{X}} \propto \dot{m}^2$ \citep{1973A&A....24..337S,1995ApJ...452..710N}, therefore, the radio/X-ray correlation can be derived, $L_{\rm{radio}} \propto L_{\rm{X}}^{0.7}$ \citep[e.g.][]{2003MNRAS.344...60G}. This model assumes that optically thick jet spectrum extends from the radio to the optical/near-infrared (OIR) band \citep[and even to the UV band, see][for more details]{2013MNRAS.429..815R}, so the relation $L_{\rm{radio}} \propto L_{\rm{UV/optical}}$ is obtained, and $L_{\rm{UV/optical}} \propto L_{\rm{X}}^{0.7}$ is theoretically expected. Recent studies have shown that there is a large population of radio-faint BH-LMXBs, some of which show a shallower radio$-$X-ray correlation at low $L_{\rm{X}}$ and a steeper one at high $L_{\rm{X}}$ \citep{2011MNRAS.414..677C, 2014MNRAS.445..290G, 2019ApJ...871...26K, 2021MNRAS.505L..58C}, which would translate to different optical$-$X-ray correlations.
 
\item The UV/optical emission is dominated by a viscously heated disc \citep{2006MNRAS.371.1334R}. In this case, the UV/optical luminosity is also linked through mass accretion rate. For a typical outer disc temperature of 8000-12000 K, the values of $\beta$ are calculated via $L_{\rm{UV/optical}} \propto \dot{m}^{\gamma}$ and $L_{X} \propto \dot{m}^2$ in hard state BH-LMXBs \citep{2002apa..book.....F,2006MNRAS.371.1334R}. The value of $\beta$ increases as the wavelength decreases: $\beta \sim 0.3$ ($\gamma \sim 0.6$) in the UV band, $\beta \sim 0.25$ ($\gamma \sim 0.5$) in the V-band, while $\beta \sim 0.15$ ($\gamma \sim 0.3$) in the K-band.
\end{enumerate}  

The BH-LMXB \source\ was discovered by the \swift\ Burst Alert Telescope ({\it BAT}) on May 30 2005 \citep{2005ATel..546....1P}. \source\ has a very short orbital period of $\sim$ 3.24 hrs \citep{2008ApJ...681.1458Z}.
\citet{2016MNRAS.463.1314S} derived a lower limit mass of the central object, $M>7.4 \pm 1.2 M_\odot$. The distance to the source is poorly constrained, $\sim2.5-8$ kpc \citep{2008ApJ...681.1458Z}. 
\citet{2019MNRAS.489.3116A} used {\em Gaia} and the LMXB Milky Way density prior to derive $8.8_{-4.0}^{+12}$ kpc. \citet{2019MNRAS.485.2642G} also reported a similar result using their Milky Way prior.
The source exhibits some outburst characteristics different from the typical LMXBs. First, after its outburst peak, the flux of \source\ began to decline exponentially, as is typical in LMXBs, however, instead of returning to quiescence, the source remained in a long-term active state. After being active for $\sim11$ years, \source\ went into quiescence in November 2016 and then underwent two mini-outbursts from late January 2017 to July 2017 \citep{2017ATel10097....1Z,2017ATel10075....1A,2017ATel10110....1B,2017ATel10325....1B,2017ApJ...848...92P,2019MNRAS.482.1840S,2019ApJ...876....5Z}. Second, \source\ has remained in the hard state for most of its outburst time, and occasionally experienced failed state transition \citep{2013MNRAS.429.1244S, 2015PASJ...67...11Y}, and only transitioned to a low-luminosity soft state in March 2015 \citep{2016MNRAS.458.1636S}. However, the source never evolved into the HSS during the outburst like other LMXBs and returned to the hard state instead.

To understand the origin of UV/optical emission in Swift J1753.5-0127, we have analysed the simultaneous UV/optical and X-ray observation with \swift. The paper is organized as follows: we describe the observations and the details of the data reduction and analysis in Section \ref{sec:observations and data reduction}.
In Section \ref{sec:results and analysis} we show the main results of this work, such as light curve, HID, UV/optical-X-ray emission correlation. In Section \ref{sec:discussion} we discuss the possible origin of UV/optical emission in \source. 
Finally, in Section \ref{sec:conclusion} we provide a summary of this work.

\section{Observations and Data Reduction}
\label{sec:observations and data reduction}
We collected all available archival \textit{Neil Gehrels Swift Observatory} \citep{2004ApJ...611.1005G} observations of J1753 from the onset of the 2005 outburst until 2017. In order to analyse the correlation between the UV/optical and X-ray emission, we have used the simultaneous X-ray and UV/optical observations from XRT and UVOT, respectively. This is the first time that this kind of correlations of the source is carried out over the 12-year long outburst.

\subsection{XRT data reduction}
\label{subsec:xrt data reduction}
We obtained all X-ray spectra and response files making use of the \swift /XRT online product builder\footnote{http://www.swift.ac.uk/user$\_$objects/} \citep{2009MNRAS.397.1177E}. Since the Photon Counting (PC) mode spectra are significantly affected by pile-up at high $L_{\rm{X}}$ and do not have enough photons to use chi-square statistics at low $L_{\rm{X}}$, we only use the data obtained in Windowed Timing (WT) mode in this paper. 
All spectra are grouped to have at least 20 counts per energy bin using the ftool \textit{grppha}. We use {\sc xspec} v.12.11.1 \citep{1996ASPC..101...17A} to analyse all X-ray spectra. We apply the empirical {\tt diskbb} and {\tt powerlaw} models to describe the disc emission and Comptonized emission, respectively. We adopt the {\tt tbabs} model and the {\tt wilm} abundance table for the interstellar absorption \citep{2000ApJ...542..914W}. We fixed the column density to $N_{\mathrm{H}} = 2.3 \times 10^{21}~\mathrm{cm}^{-2}$ \citep{2006ApJ...652L.113M}.
We have used two models to fit each spectrum, e.g. an absorbed power-law {\tt tbabs*(powerlaw)}, and an absorbed disc-blackbody plus a power-law {\tt tbabs*(diskbb+powerlaw)}. We computed each fitting parameter with $1\sigma$ error using the {\tt err} command. The best-fitting model is determined by an F-test.

Once the best-fitting model had been determined, we calculated the unabsorbed X-ray fluxes in the 0.3$-$2 keV, 2$-$10 keV, and 0.3$-$10 keV range, respectively, using the {\tt cflux} convolution model. Errors on the unabsorbed X-ray fluxes are determined at $1\sigma$ confidence level.

\subsection{UVOT data reduction}
\label{subsec:uvot data reduction}
\swift /UVOT has six filters from the ultraviolet to the optical band. The effective wavelength ($\lambda_{\rm eff}$) of each filter are \citep[][]{2008MNRAS.383..627P}: {\it UVW2 (2030\AA)}, {\it UVM2 (2231\AA)}, {\it UVW1 (2634\AA)}, {\it U (3501\AA)}, {\it B (4329\AA)}, {\it V (5402\AA)}. These wavelength values are used in the following extinction correction process. All the UVOT observations were taken in image mode with one or more filters. The UVOT data is processed using {\sc heasoft} v.6.28. We use the {\tt uvotimsum} tool to sum the sky images in observations with more than one image extension. We determine the flux density of each observation with the {\tt uvotsource} tool. We selected a circular region with a radius of 5 arcsec centred on the source, and a circular source-free region with a radius of 20 arcsec for the background correction. Errors on the UV/optical flux densities are determined at $1\sigma$ confidence level.

In order to study the correlation between the UV/optical and the X-ray emission, we converted the flux densities ($F_{\lambda}$; $erg\ s^{-1}\ cm^{-2}\ $\AA$^{-1}$) to fluxes ($F$; $erg\ s^{-1}\ cm^{-2}$) by multiplying the full-width at half-maximum \citep[FWHM, $\Delta\lambda$; from][]{2008MNRAS.383..627P} of the filter bandpass (i.e. $F \approx \Delta\lambda F_{\lambda}$).

We deredden the UV/optical fluxes for Galactic extinction making use of the \citet{1999PASP..111...63F} parameterization in each filter. We used the color excess E(B-V)=0.45 \citep{2014ApJ...780...48F, 2015ApJ...810..161R}, and adopted the galactic extinction law with $R_V=A_V/E(B-V)=3.1$.
Once this two parameters was determined, we use the {\it unred}\footnote{https://pyastronomy.readthedocs.io/en/latest/pyaslDoc/aslDoc/\\unredDoc.html}
python modular in PyAstronomy based on \citet{1999PASP..111...63F} parameterization to deredden the UV/optical data.

\begin{figure}
\begin{center}
\includegraphics[scale=.33]{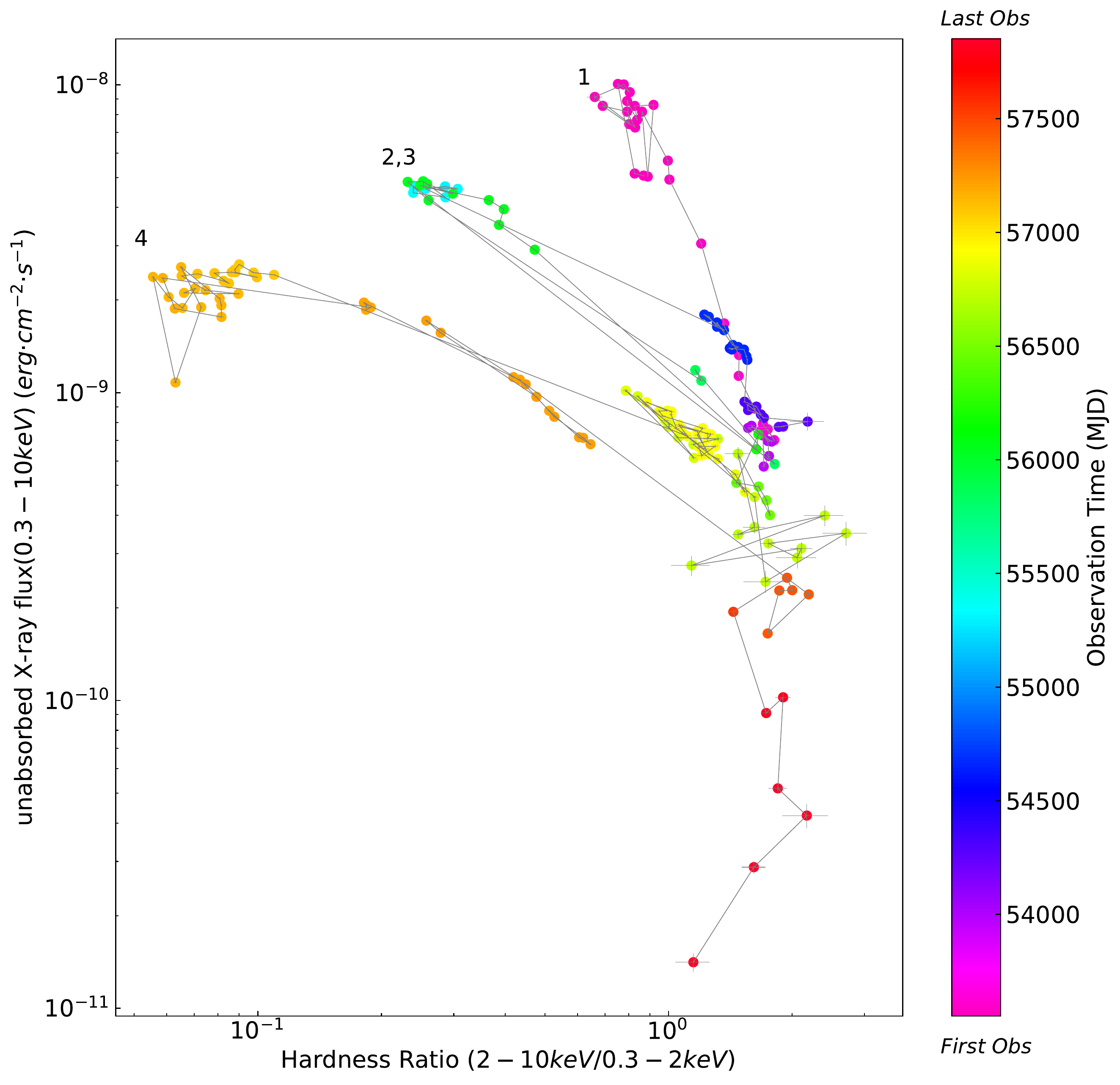}
\caption{Hardness-intensity diagram of \source. The color scale represents the time of observation. Marks 1,2,3 and 4 represent, respectively, the location of the main outburst and several low-intensity peaks in Figure \ref{fig:para_time}.}
\label{fig:HID_time}
\end{center}
\end{figure}

\section{Results and Analysis}
\label{sec:results and analysis}
\subsection{UV/Optical and X-ray light curves}
\label{subsec:lc}
The $\sim$12 years unabsorbed X-ray (0.3$-$10 keV) and UV/optical light curves of J1753 are shown in the first and second panel of Figure \ref{fig:para_time}, respectively. The X-ray light curve shows the typical exponential decay at the beginning of outburst, followed by an unusual long-term low-level activity with several low-intensity peaks (the 2,3,4 gray areas in Figure \ref{fig:para_time}). The UV/optical light curve also displays the same characteristic decay at the early outburst, however, the later low-level activity are not as obvious as that in the X-ray band. 
In panels (d), (e) and (f) of Figure \ref{fig:para_time} we show the best-fitting X-ray spectral photon index, temperature at inner disc radius and the hardness ratio (HR), respectively. We note that the disc component is not always needed in all observations. From the variation of the photon index with time, we know that J1753 stayed in the hard state for a long time. In Figure \ref{fig:para_time}, we mark five gray shadow regions. The first gray area is the main outburst, and the middle three gray areas are the low-intensity peaks with HR <0.7, the last gray area is the mini-outburst after the first quiescence. We note that the gray areas labeled 2 and 3 correspond to the failed state transition \citep{2013MNRAS.429.1244S, 2013MNRAS.433..740S}, the third gray area also is the second BAT flux "dip" from \citet{2013MNRAS.433..740S}, and the fourth gray area is the low-luminosity soft state from \citet{2016MNRAS.458.1636S}. In the long period of low-level active state, at each low-intensity peak, the corresponding photon index is larger than 2.0, inner disc temperature are larger than 0.3, and the hardness ratios are less than 0.7.


\begin{figure}
\begin{center}
\includegraphics[scale=.6]{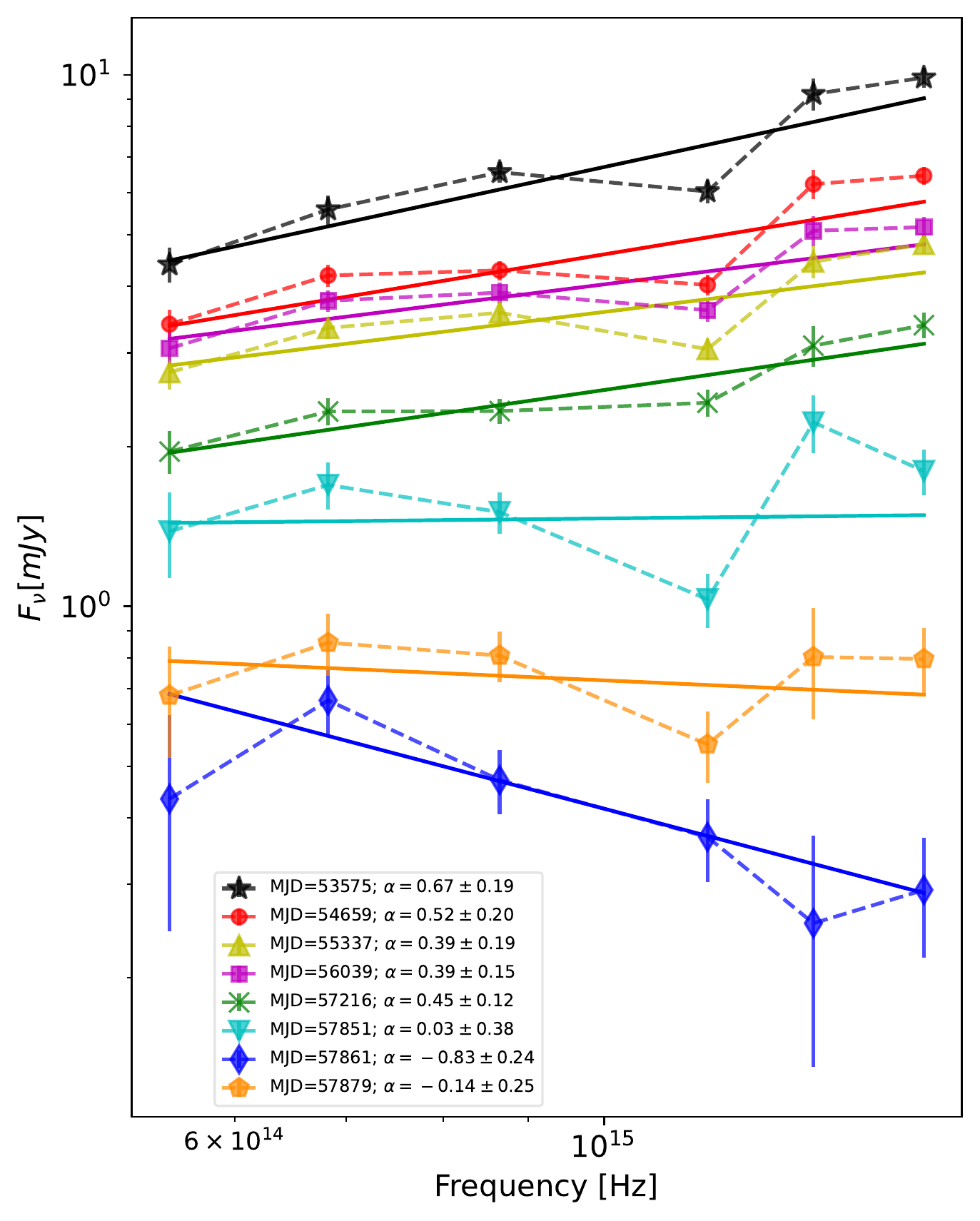}
\caption{Representative UVOT SED plot of \source\ fitted by a simple power-law model.}
\label{fig:sed}
\end{center}
\end{figure}
\begin{figure*}
\begin{center}
\includegraphics[scale=.35]{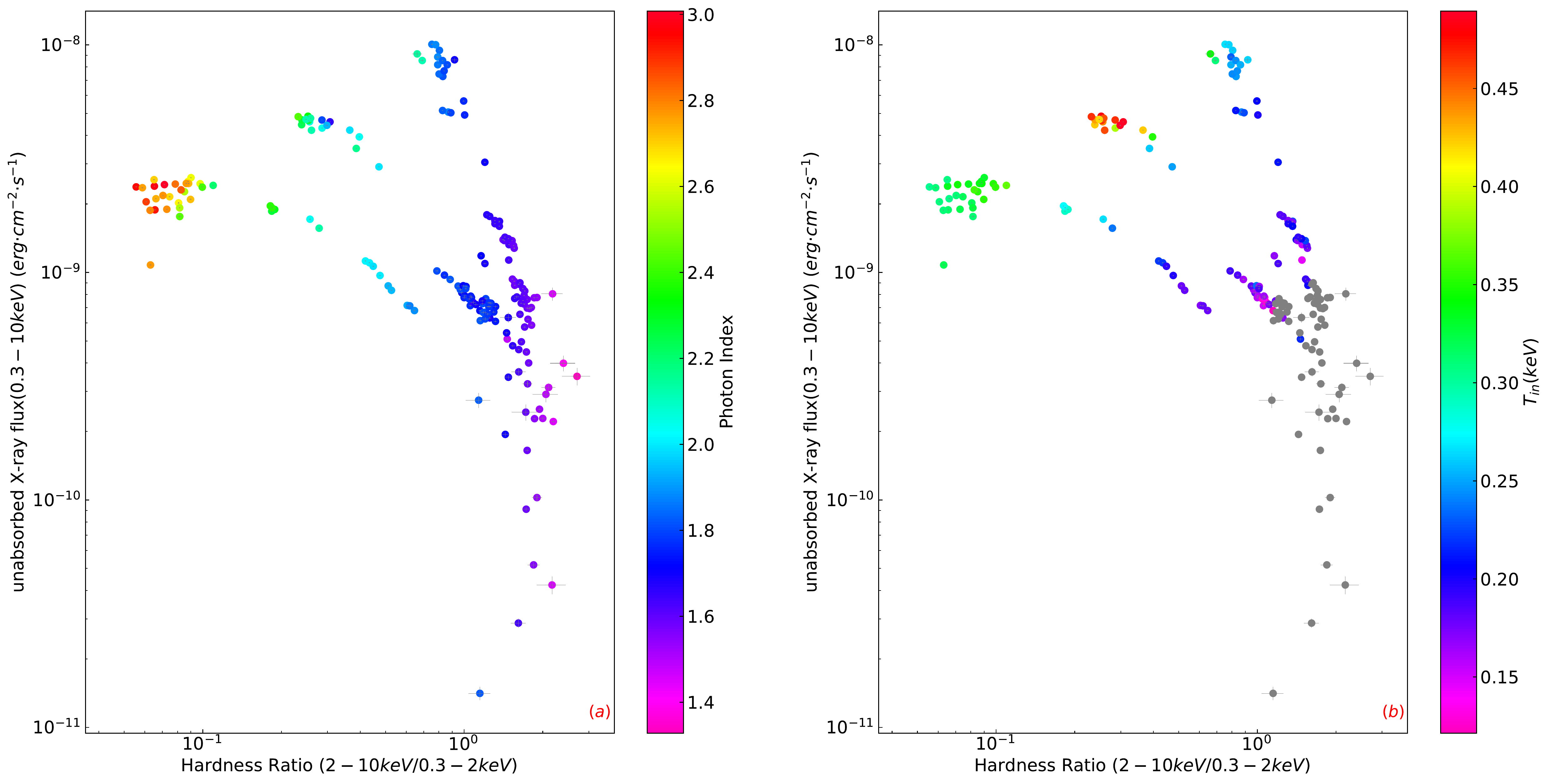}
\caption{The spectral evolution of \source\ in the HID. The colors of the points represent photon index (panel a) and inner disc temperature (panel b). The gray points represent the data points with no "diskbb" component.}
\label{fig:HID_index_tin}
\end{center}
\end{figure*}
\begin{figure*}
\begin{center}
\includegraphics[scale=.36]{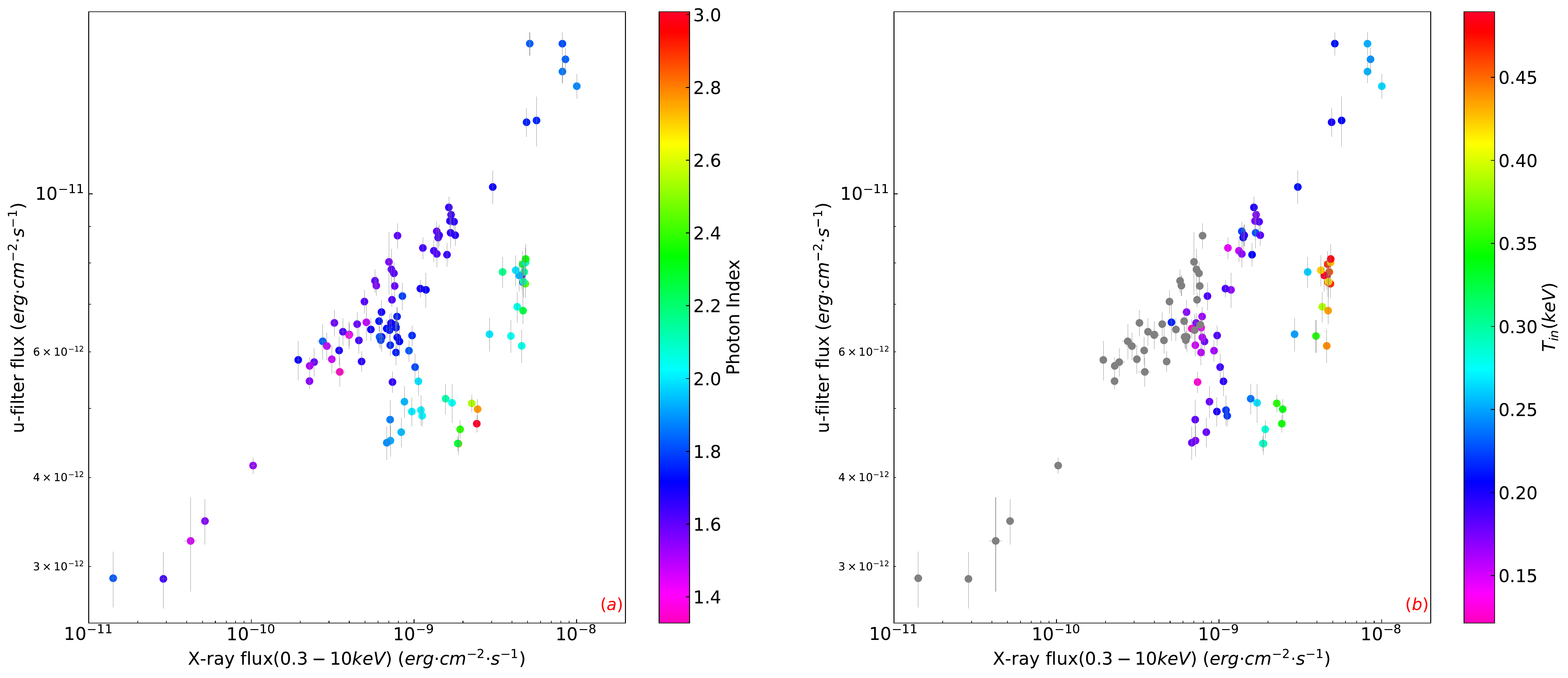}
\caption{The spectral evolution of \source\ in U filter vs. X-ray flux diagram. The colors of the points represent photon index (panel a) and inner disc temperature (panel b). The gray points represent the data points with no \texttt{diskbb} component.}
\label{fig:u-x_correlation_index_tin}
\end{center}
\end{figure*}
\begin{figure*}
\begin{center}
\includegraphics[scale=.60]{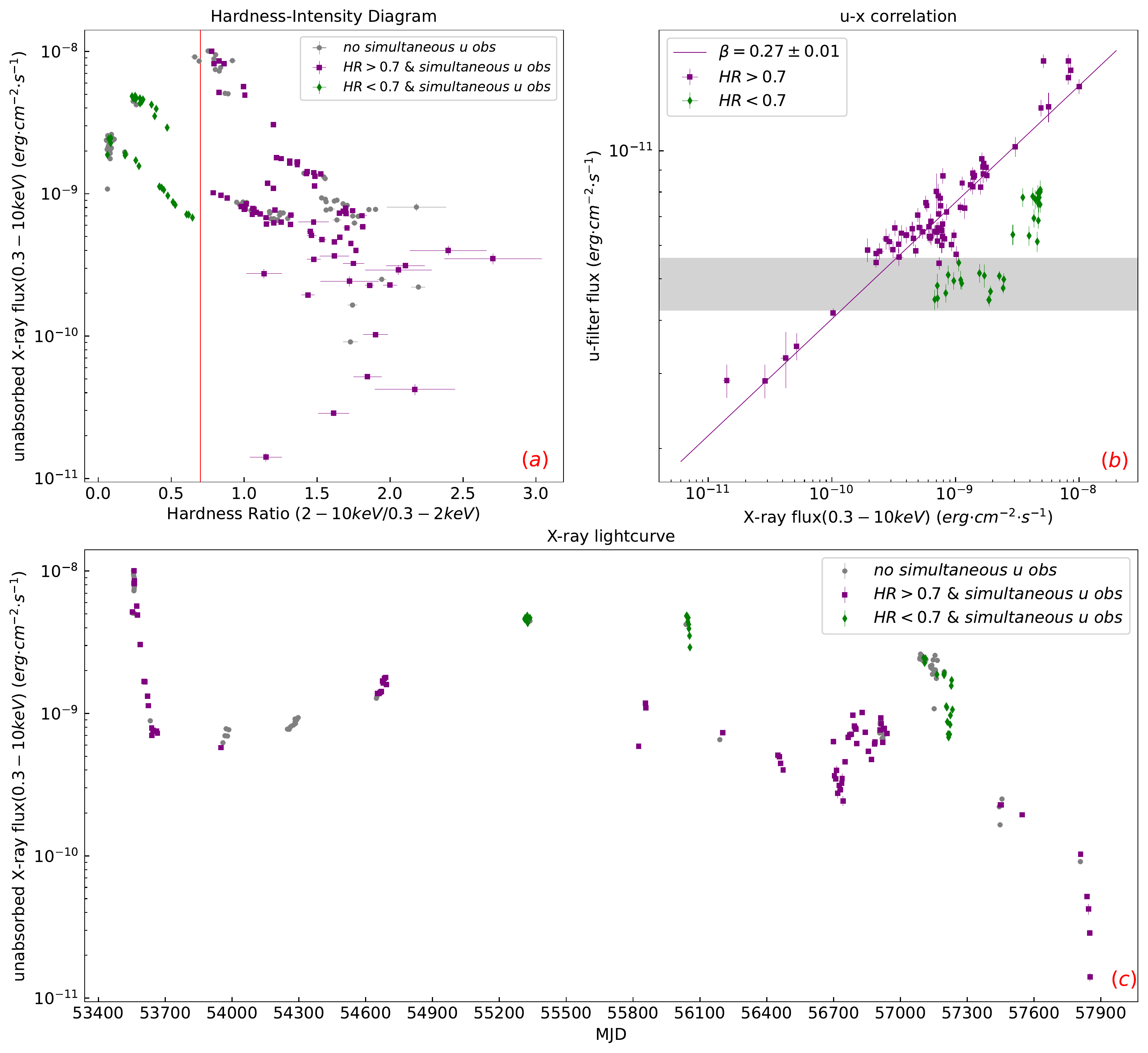}
\caption{The correlation analysis taking U-X as an example. These purple squares represent data points with HR > 0.7 and simultaneous observation in the U filter, green diamonds represent data points with HR < 0.7 and simultaneous observation in the U filter, gray points represent data points with no simultaneous observation in U filter. Panel (a): HID, the solid red line represents the dividing line of HR=0.7. Panel (b): dereddened U filter fluxes versus unabsorbed X-ray (0.3$-$10 keV) flux, the gray shadowed area represent the observation of the 4th low-intensity peak. The purple solid line is a power-law fit to purple square (HR > 0.7) data set. Panel (c): X-ray (0.3$-$10 keV) light curve.} 
\label{fig:u_x_HID_correlation_lc}
\end{center}
\end{figure*}

\subsection{Spectral index of UV/Optical SED}
\label{subsec:sed}

During the outburst, the source was observed simultaneously with six UVOT filters in many occasions. We make the spectral energy distribution (SED) between the optical and ultraviolet bands at different time periods and fitted them with a power-law function, $S_{\nu} \propto {\nu}^{\alpha}$, where $S_{\nu}$ is the dereddened UV/optical flux density at a certain frequency $\nu$ and $\alpha$ is the spectral index of the SED, we only used six filters to measure $\alpha$. Figure \ref{fig:sed} shows some fitted UVOT SED plots at representative periods (see arrows in Figure \ref{fig:para_time}-(c)). The SED were fitted by the \texttt{curve\_fit}\footnote{https://docs.scipy.org/doc/scipy/reference/generated\\/scipy.optimize.curve\_fit.html} function from python package \texttt{scipy.optimize}. The variation of the UV/optical spectral index with the observation time is shown in panel (c) of Figure \ref{fig:para_time}. During the outburst before MJD $\sim 57750$, most of the UV/optical spectral index are positive around $\alpha \sim 0.4 \pm 0.1$. At the end of the outburst, as the UV/optical flux decreases, the values of $\alpha$ turn into negative, which indicate that the UV/optical spectral shape have changed significantly during the outburst decay phase, which is similar to what found in \citet{2019ApJ...876....5Z}.

\subsection{Hardness-Intensity Diagram}
\label{subsec:hid}
In Figure \ref{fig:HID_time}, we plot the HID of J1753, in which the hardness ratio is defined as the ratio of the unabsorbed X-ray flux in the 2$-$10 keV and 0.3$-$2 keV bands, and the intensity is defined as the unabsorbed X-ray flux in the 0.3$-$10 keV band. The symbols with the color from purple to red in the figure display the evolution of HR and X-ray flux with observation time from the start to the end of the outburst. The tracks in the HID can be simply viewed as several branches, which represent the early stage of the outburst and several low-intensity peaks, respectively. For several branches, the data with higher peak flux have correspondingly softer spectra.

\subsubsection{$\Gamma$ and $T_{in}$ in HID}
\label{subsec:Gamma_tin_hid}

We also plot the distribution of the X-ray photon index ($\Gamma$) and inner disc temperature ($T_{in}$) in the HID (see Figure \ref{fig:HID_index_tin}). In the panel (a) of Figure \ref{fig:HID_index_tin}, the upper left part shows that the X-ray spectral photon index of several branches increases as the HR and the low-intensity peak's flux decrease and the corresponding peak's photon index increases (also see panel (d) of Figure \ref{fig:para_time}). In the panel (b) of Figure \ref{fig:HID_index_tin}, the \texttt{diskbb} component mainly appears in the observations of the upper left branches.

\subsection{The X-ray and UV/optical Flux Correlations}
\label{subsec:x_uv/opt correlation}
As mentioned in Section \ref{subsec:hid}, the hardness ratio always decreases during the low-intensity peaks in the long outburst. In order to facilitate the analysis of the correlation between optical and X-ray emission, we divide all the data into two parts, as shown in panel (a) of Figure \ref{fig:u_x_HID_correlation_lc}. From the Figure \ref{fig:para_time}, especially the panel (f), and the description in the Section \ref{subsec:lc}, the HR > 0.7 boundary is a clear cutoff that separates the two "failed-transition" states  and the low-luminosity soft from the hard state.
We then reflect these two parts in the correlation distribution map. Panel (b) of Figure \ref{fig:u_x_HID_correlation_lc} shows the U filter fluxes versus the 0.3$-$10 keV X-ray fluxes. It is apparent from this figure that most of the data with HR > 0.7 show a good correlation between U band and X-ray fluxes. We also plot other UVOT filter fluxes against the 0.3$-$10 keV X-ray fluxes in Figure \ref{fig:other filter vs 0.3-10 X} of Appendix A.

For the observations with HR > 0.7, we first perform a Spearman's rank test and find that there is an apparent correlation between the UV/optical and X-ray fluxes. In the last two columns of Table \ref{table:fit results} we show the Spearman's correlation coefficient ($\rho$) and  the null-hypothesis probability of the Spearman's correlation test ($P$), respectively. The results of the Spearman's rank test indicate that there is a strong correlation between UV/optical and X-ray fluxes in the data points with HR > 0.7. We fit a power-law function, $F_{\rm{UV/optical}} = \alpha F_{X}^{\beta}$, to the UV/optical and 0.3$-$10 keV X-ray fluxes. 
The best fitting $\alpha$ and $\beta$ values are given in Table \ref{table:fit results}. We find that as the effective wavelength ($\lambda_{\rm eff}$) of the UVOT filters decreases from $\sim$ 5402 \AA\ to $\sim$ 2030 \AA, the best fit power-law index ($\beta$) increases from $\sim0.24$ to $\sim0.33$.

In previous correlation studies, the X-ray flux between 2 and 10 keV bands was usually calculated \citep[e.g.][]{2006MNRAS.371.1334R, 2013MNRAS.428.3083A, 2015MNRAS.453.3461S, 2020MNRAS.493..940L, 2020arXiv201200169Y}. In order to compare with previous work,
we carry out the same procedure using the 2$-$10 keV X-ray flux. The correlations between UV/optical and 2$-$10 keV X-ray fluxes are shown in the Figure \ref{fig:uvot filter vs 2-10 X} of Appendix A and the panel (b) of Figure \ref{fig:uvot-X_correlation}. Table \ref{table:fit results} shows the fitting results for each correlation using the 2$-$10 keV X-ray flux. In this case the power-law index ($\beta$) increases from $\sim0.26$ to $\sim0.37$ when the effective wavelength($\lambda_{\rm eff}$) of the UVOT filters decreases from 5402 \AA\ to 2030 \AA. For comparison purposes, we also plotted the correlation  between UV/optical and the two X-ray bands in the same frame in Figure \ref{fig:x_two_band_correlation} of Appendix A.

\subsubsection{$\Gamma$ and $T_{\rm in}$  in correlation diagram}
\label{subsec:Gamma_tin_correlation}
In order to understand the evolution of the X-ray spectral parameters during the outburst, we plot the data with the color of the markers on the basis of $\Gamma$ and $T_{in}$ in the UV/optical$-$X-ray correlation diagram. In the panel (a) of Figure \ref{fig:u-x_correlation_index_tin}, we show the distribution of photon index in the UV/optical$-$X-ray correlation diagram (taking U filter as an example). In the panel (b), we show the distribution of inner disc temperature in the U$-$X-ray correlation diagram. The gray points represent the data points with no \texttt{diskbb} component. As is evident from Figure \ref{fig:u-x_correlation_index_tin}, the data with high photon index and high inner disc temperature deviate from the UV/optical$-$X-ray correlation.

\begin{figure*}
\begin{center}
\subfigure{\includegraphics[width=0.96\columnwidth]{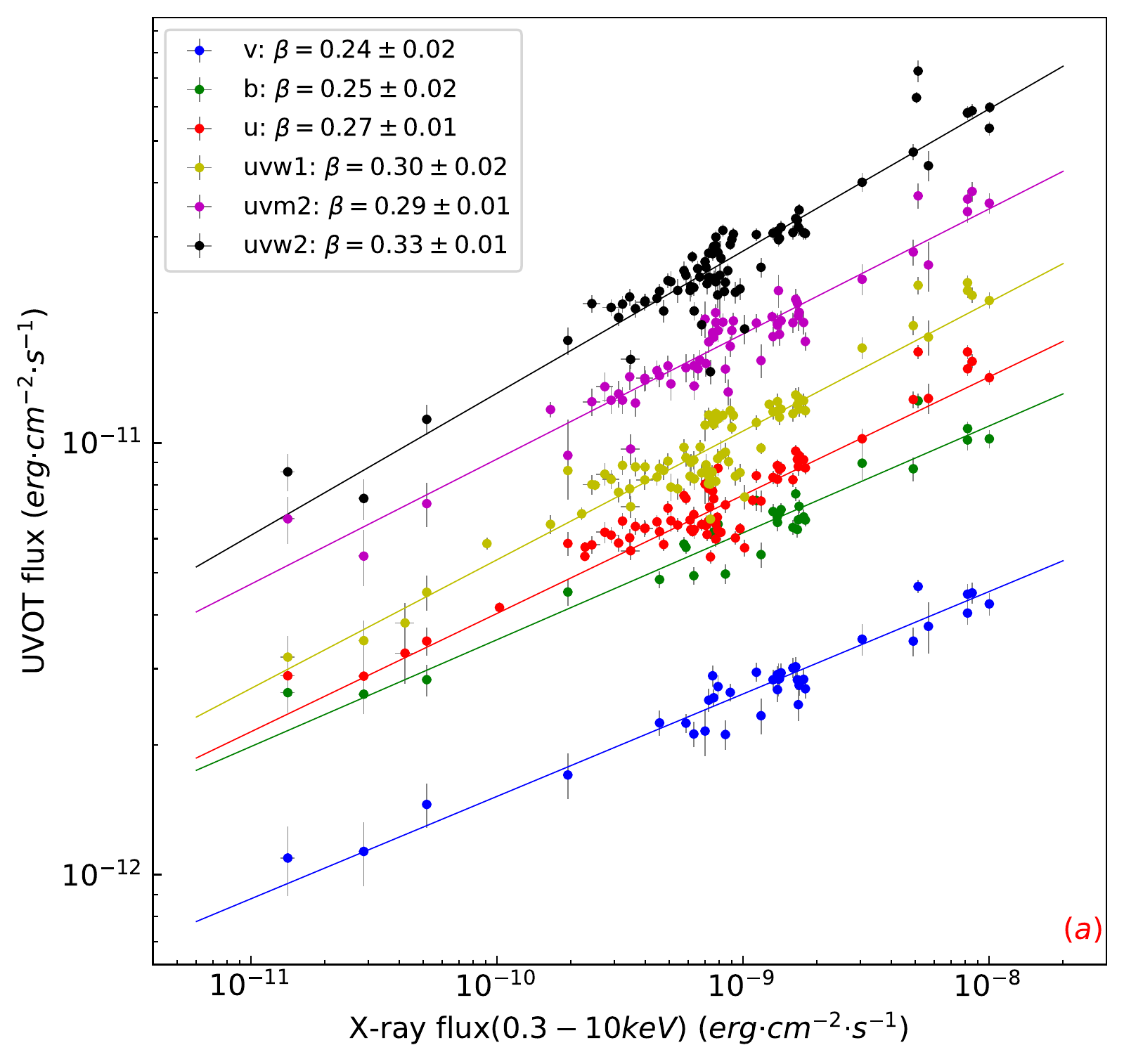}}
\subfigure{\includegraphics[width=0.96\columnwidth]{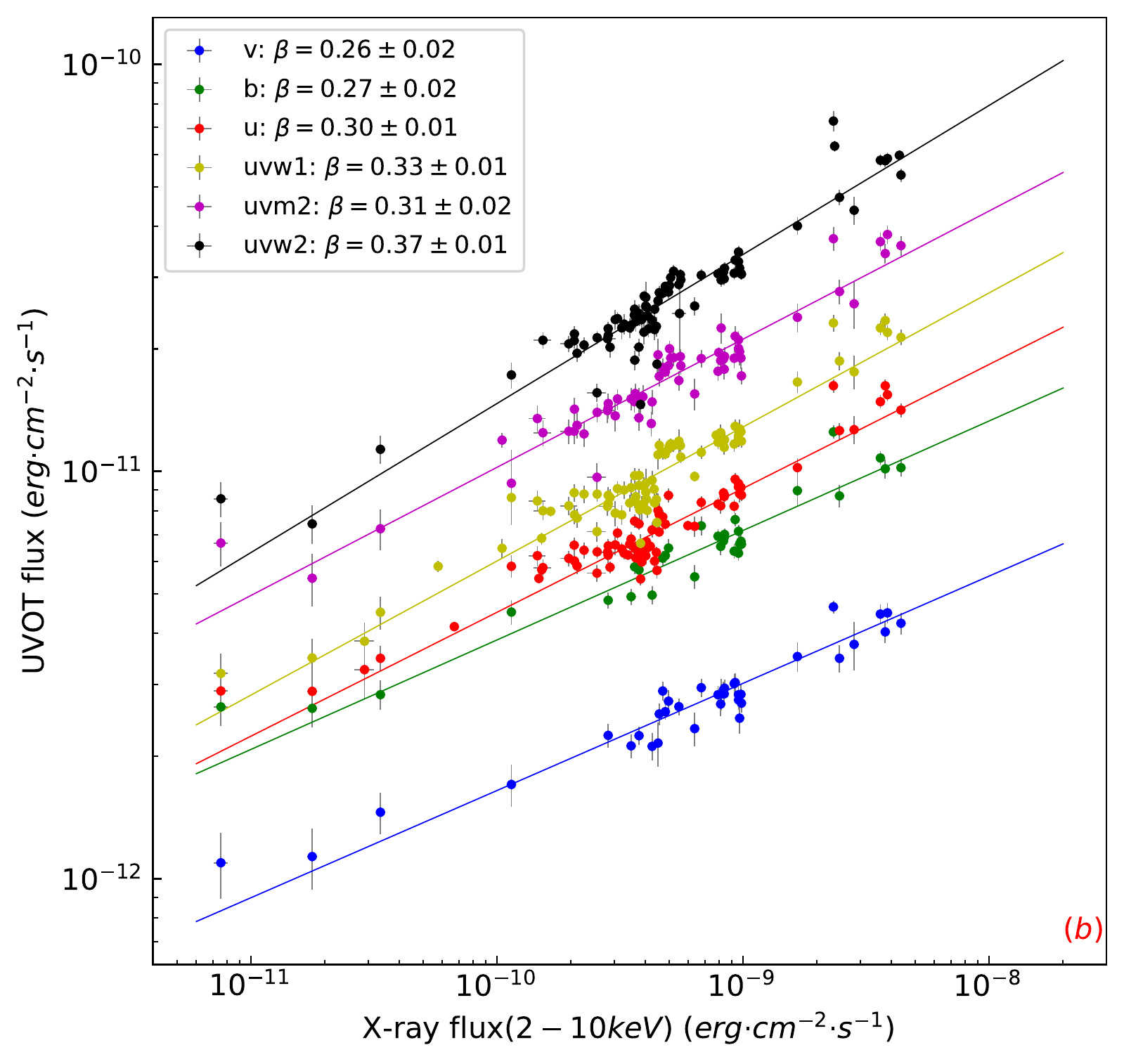}}
\caption{The UVOT-X-ray correlation of \source\ in HR>0.7 observations. panel(a): correlation between the dereddened UV/optical fluxes and the unabsorbed X-ray (0.3$-$10 keV) flux.panel(b): correlation between the dereddened UV/optical fluxes and the unabsorbed X-ray (2$-$10 keV) flux. Only the data points with HR > 0.7 are fitted in both figures.}
\label{fig:uvot-X_correlation}
\end{center}
\end{figure*}
\begin{table*}
\begin{center}
\caption{Results of Spearman's rank test and power-law fit ($F_{\rm{UV/optical}} = \alpha F_{X}^{\beta}$) to the UV/optical and the X-ray (0.3$-$10 keV, 2$-$10 keV) correlation.}\label{table:fit results}
\begin{threeparttable}
\begin{tabular}{cccccccccc}\hline\hline
Filter & $\lambda_{\rm eff}$ & \multicolumn{4}{c}{0.3$-$10 keV} & \multicolumn{4}{c}{2$-$10 keV} \\
\cmidrule(r){3-6} \cmidrule(r){7-10}
 & (\AA) & $\beta \pm \Delta \beta$ & $\alpha \pm \Delta \alpha$ & $\rho$\tnote{*} & $P$\tnote{**}
 & $\beta \pm \Delta \beta$ & $\alpha \pm \Delta \alpha$ & $\rho$\tnote{*} & $P$\tnote{**} \\
\hline
$v$ & 5402 & $0.24\pm0.02$ & $3.57(\pm 1.13)\times10^{-10}$ & 0.85 & $2.19\times10^{-11}$ & $0.26\pm0.02$ & $7.07(\pm 2.65)\times10^{-10}$ & 0.87 & $3.73\times10^{-12}$ \\
$b$ & 4329 & $0.25\pm0.02$ & $1.04(\pm 0.38)\times10^{-9}$ & 0.88 & $2.69\times10^{-11}$ & $0.27\pm0.02$ & $1.88(\pm 0.79)\times10^{-9}$ & 0.89 & $6.09\times10^{-12}$ \\
$u$ & 3501 & $0.27\pm0.01$ & $2.20(\pm 0.65)\times10^{-9}$ & 0.82 & $1.62\times10^{-19}$ & $0.30\pm0.01$ & $4.98(\pm 1.40)\times10^{-9}$ & 0.87 & $1.15\times10^{-24}$ \\
$uvw1$ & 2634 & $0.30\pm0.02$ & $5.12(\pm 1.63)\times10^{-9}$ & 0.84 & $3.20\times10^{-23}$ & $0.33\pm0.01$ & $1.18(\pm 0.37)\times10^{-8}$ & 0.89 & $7.40\times10^{-29}$ \\
$uvm2$ & 2231 & $0.29\pm0.01$ & $7.20(\pm 2.12)\times10^{-9}$ & 0.91 & $4.58\times10^{-24}$ & $0.31\pm0.02$ & $1.44(\pm 0.48)\times10^{-8}$ & 0.93 & $4.63\times10^{-27}$ \\
$uvw2$ & 2030 & $0.33\pm0.01$ & $2.54(\pm 0.74)\times10^{-8}$ & 0.85 & $3.20\times10^{-24}$ & $0.37\pm0.01$ & $6.73(\pm 2.05)\times10^{-8}$ & 0.92 & $1.24\times10^{-33}$ \\
\hline
\end{tabular}
\begin{tablenotes}
 \item[*] Spearman's correlation coefficient, which varies between -1 and +1 with 0 implying no correlation and $\rho = +1$ or $\rho = -1$ implying, respectively, positive or negtive correlation.
\item[**] Null hypothesis probability of Spearman's correlation test. The P-value indicates the probability of an uncorrelated system producing datasets that have a Spearman correlation.
\end{tablenotes}
\end{threeparttable}
\end{center}
\end{table*}

\begin{figure}
\begin{center}
\includegraphics[scale=.55]{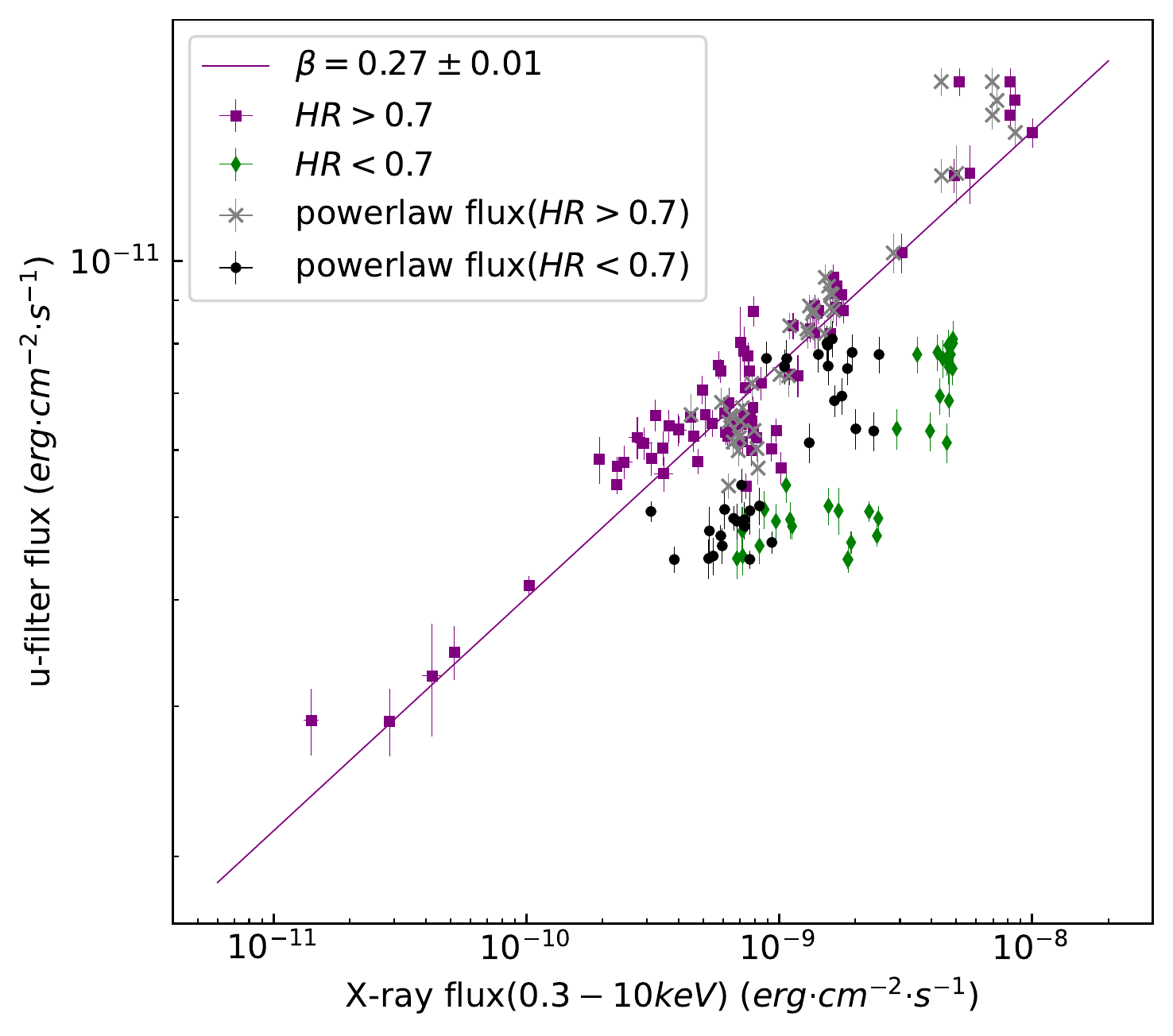}
\caption{U filter-X-ray (0.3$-$10 keV) flux correlation after subtracting the X-ray flux from {\tt diskbb}. The purple squares and green diamonds represent data with HR>0.7 and HR<0.7 before subtracting the {\tt diskbb} flux, respectively. The gray crosses and black filled circles represent data with HR>0.7 and HR<0.7 after subtracting the {\tt diskbb} flux, respectively. The purple solid line is a power-law fit to the purple points.}
\label{fig:changes_in_u_x_correlation}
\end{center}
\end{figure}

\section{Discussion}
\label{sec:discussion}
We have analysed all available \swift/XRT and \swift/UVOT data of the BH-LMXB \source. We have studied the UV/optical and X-ray flux correlations during the whole 12-year long-term outburst, and found, for the first time, that the power-law index $\beta$ increases from $\sim$ 0.24 to $\sim$ 0.33 as the UV/optical wavelength decreases from $\sim$ 5400 \AA\ (V) to $\sim$ 2030 \AA\ (UVW2). We found that the data that deviate from the correlation are the low-intensity peaks that appear in the X-ray band during the outburst, and found these low-intensity peaks were dominated by the soft X-ray component.

\subsection{Several low-intensity peaks}
\label{subsec:low-intensity peaks}
As shown in Figure \ref{fig:para_time}, the last three gray shadowed areas represent the time intervals when three low-intensity peaks appear. We find that the source spectra become soft in these three periods. In particular, during the fourth peak, the source went into a low-luminosity soft state \citep{2016MNRAS.458.1636S}. As mentioned in Section \ref{subsec:lc}, in these three periods the disc component appeared, and the X-ray photon index was significantly larger than that in the rest of the outburst, which indicates that the soft X-ray emission increased.
In order to estimate the soft X-ray contribution during these three low-intensity peaks, we calculated the total unabsorbed X-ray flux and the {\tt diskbb}-subtracted X-ray flux between 0.3 keV and 10 keV. In Figure \ref{fig:changes_of_corr_after_del_disc_flux} and Figure \ref{fig:changes_in_u_x_correlation} we show the total (0.3$-$10 keV) and  the {\tt diskbb}-subtracted X-ray flux  during these three peaks {for} HR < 0.7. We also did the same calculation for the data with HR > 0.7. We find that the X-ray flux is significantly reduced after removing the disc component during these low-intensity peaks when HR < 0.7, whereas the X-ray flux does not show any significant reduction after subtracting the disc component in the data with HR > 0.7. This indicates that the data during these peaks that deviate from the UV/optical$-$X-ray correlation are strongly affected by the soft X-ray emission from inner disc. 
Moreover, previous study showed a reduction of the optical emission during the fourth peak \citep{2019ApJ...876....5Z}. Therefore the deviations from the UV/optical$-$X-ray correlation due to the combination of both an increase of the soft X-ray emission and reduction of the UV/optical emission. Since the variation of the UV/optical flux is comparatively small, we think that the deviation from the correlation is mostly driven by an increase of the soft X-ray flux in the inner part of the accretion disc.

We also find that, taking the U filter as an example, in the gray shaded area of Panel (b) in Figure \ref{fig:u_x_HID_correlation_lc}, the U/optical flux during the fourth peak changes slightly, whereas the X-ray flux varies by a factor of four. This indicates that the UV/optical flux does not strongly affected by the X-ray emission during the fourth low-intensity peak. Similar trends can not be excluded in the  UV/optical$-$X-ray diagram for both peak 2 and 3 since there is no good \swift\ UV/optical data coverage during peak 2 and 3 as there is for peak 4. Furthermore, given the smooth evolution of the UV/optical light curves in Figure 2 of \citet{2019ApJ...876....5Z} and similar spectral properties shown in Figure \ref{fig:u-x_correlation_index_tin}, we speculate that the UV/optical flux neither in peak 2 nor in peak 3 is correlated with the X-ray flux. In addition, if the X-ray reprocessing dominates the UV/optical emission, the UV/optical band should have the corresponding increases during these three X-ray peaks. We therefore suggest that the X-ray flux does not strongly affect the UV/optical flux.

\subsection{UV/optical spectral index}
\label{subsec:spectral_index}

In the above analysis we find that in the long-term outburst the UV/optical light curve did not show significant variations, and the UV/optical spectral index (within errors) stayed almost constant before the source went into quiescence. The UV/optical spectra were not strongly affected by the increased X-ray flux during the three low-intensity peaks.

\citet{2006MNRAS.371.1334R} studied  the optical/near-infrared SEDs of 15 BH-LMXBs in which the spectral index is almost positive, and they suggested that these spectra could have a thermal origin. In our analysis, the UV/optical spectral index varied around $ \sim 0.4$ before the source entered into the outburst decay phase. The UV-optical spectral index is $\sim 0.4 $ before the outburst decay, whereas the spectral index between the $V$ and $i'$ bands is $\sim 0.9 $ \citep{2019ApJ...876....5Z}. This implies a curvature, turning down at longer wavelengths, which is consistent with the accretion disc approaching the Rayleigh-Jeans tail at longer wavelengths. The values of the UV/optical spectral indices evolved from positive to negative during the period when the source went into quiescence, which indicates that the temperature of the disc decreases at large radii, consistent with the findings in \citet[][in particular Figure 4]{2019ApJ...876....5Z} and can be explained by a cooling disc. Previous studies of broadband SED \citep[e.g.][]{2014ApJ...780...48F,2015ApJ...810..161R,2015ApJ...808...85T,2019MNRAS.482.1840S} suggested that \source\ had a cool disc at large radii during the outburst. Assuming the UV/optical emission comes from the thermal emission of the cool disc, the almost constant spectral index indicates that the disc component at large radii evolved very slowly and was not significantly affected by the rapid activities (i.e. several low-intensity X-ray peaks) from the inner region of the accretion disc. 

\subsection{UV/optical-X-ray correlation}
\label{subsec:correlation}
In BH-LMXBs, we expect different UV/optical emission processes to
yield different correlations with the X-ray flux. Studying the correlations between the emission in different wavelengths can provide valuable information about the accretion process and any associated outflows. The power-law index of the relation between the UV/optical and X-ray fluxes is often used to identify the origin of the UV/optical emission mechanism. 

If the UV/optical emission is due to the reprocessing of X-rays in the outer accretion disc, the power-law index is expected to vary from $\sim0.3$ to $\sim0.9$ as the energy band changes from optical to UV. For the X-ray reprocessing model, \citet{1994A&A...290..133V} derived a power-law index $\beta \sim0.5$ in V-band. There are some observations consistent with this\citep[e.g.][]{2016ApJ...826..149B}.
Under the assumption that the emission from an irradiated disc can extend from optical to UV, and the $\beta \sim0.5$ could be even applied to near-ultraviolet (NUV) and UV bands \citep[e.g.][]{2007ApJ...666.1129R, 2010ApJ...719.1993R}. \citet{2015MNRAS.453.3461S} suggested the value of $\beta$ should vary from optical to UV. 

If the UV/optical emission comes from synchrotron emission in jet of BH-LMXBs, the power-law index should be around $\sim0.7$ over the entire UV/optical bands \citep{2006MNRAS.371.1334R,2012MNRAS.419.1740R}. This kind of situation usually happens in the hard state of BH-LMXBs because the jet would be quenched in the soft state.
\citet{2012MNRAS.419.1740R} investigated the optical excess of the BHXB XTE J1752-223, they obtained a relation, $F_{\rm{optical}} \propto F_{X}^{0.72\pm 0.12}$, and they suggested that the origin of the optical excess is the synchrotron jet. \citet{2010MNRAS.406.1471S} obtained the spectral indices of the radio SED of \source\ are negative, between July 2005 and July 2007, and they found the jet is too weak to be responsible for the optical/infrared emission and they suggested that the thermal emission from the accretion disc dominate the optical emission.

If the UV/optical emission is due to the viscously heated disc, for BH-LMXBs, the expected power-law power-law index varies from $\sim0.15$ to $\sim0.3$ from optical to UV \citep{2002apa..book.....F,2006MNRAS.371.1334R}.
As shown in Table \ref{table:fit results}, we found that the UV/optical and X-ray fluxes are strongly correlated during the hard state. The values of the power-law indices for the UV/optical-X-ray correlation vary in the range 0.24 < $\beta$ < 0.33 and 0.26 < $\beta$ < 0.37, if we use the 0.3$-$10 keV and 2$-$10 keV X-ray bands, respectively. Compared with the three models discussed above, we find that the values of UV/optical$-$X-ray power-law index $\beta$ deviate greatly from the X-ray reprocessing model and the jet model. Our results are consistent with the model in which the UV/optical emission comes from a viscously heated disc around a black hole in the hard state. \citet{2002apa..book.....F} predicted that the power-law index is wavelength dependent. For the UV/optical emission of a viscously heated steady-state disc, as the wavelength of the UV/optical band decreases, the value of $\beta$ increases. Our results are consistent with the theoretical trend (see the values of $\beta$ in Table \ref{table:fit results}). \citet{2013MNRAS.428.3083A} and \citet{2019MNRAS.485.3064B} got similar results in studying the BH-LMXB Swift J1357.2–0933, and they suggested that the UV/optical emission during the outbursts of the source in 2011 and 2017, respectively, could be explained by the viscously heated disc. We find that there are some similarities between Swift J1357.2-0933 and \source . Both sources have short orbital periods, $\sim3$ hrs, and stayed in the hard state during outburst. However, the duration of the two outbursts of Swift J1357.2-0933 ($\sim7$ and $\sim4.5$ months, respectively) are shorter than that of the unusual long-term outburst in \source.

It has been suggested that the accretion disc of \source\ should be irradiated and heated by the central X-ray emission during the outburst \citep{2019MNRAS.482.1840S, 2019ApJ...876....5Z}. By modelling the bolometric X-ray light curve using the shape predicted by the irradiated DIM (IDIM), and fitting the UV/optical/NIR SEDs during the standstill phase and mini-outburst with an irradiated disc model, \citet{2019MNRAS.482.1840S} found a fully irradiated disc exists in the long-term outburst. \citet{2019MNRAS.482.1840S} also analysed the UV/optical$-$X-ray correlation for the mini-outburst, and the correlations they obtained are shallower than expected from X-ray reprocessing. They suggest that the UV/optical emission is a combination of X-ray reprocessing and a contribution of synchrotron emission from the corona, resulting in shallower correlations.

As discussed in section \ref{subsec:low-intensity peaks}, the low-intensity peaks only appear in the X-ray band and are dominated by the soft X-ray component. There is no significant increase in the UV/optical band during these peaks. This indicates that a high fraction of the X-ray flux is dominated by emission from the inner region of the accretion disc, and could be due to the irradiated disc instability model. Most of the optical emission is coming from the outer viscously heated disc and the disc is not irradiated enough. This is consistent with the results in \citet{2021MNRAS.507.5507A} who suggested that insufficient irradiation would occur in the so called failed-transition outburst, like in Swift J1753.5-0127. We can not exclude, however, that there is a small fraction of UV/optical emission from irradiation due to the existence of the irradiated disc. There is possibly some small contribution from the jet or the hot flow as well \citep{2008ApJ...682L..45D, 2009MNRAS.399..281H, 2014ApJ...788..184W, 2015ApJ...810..161R, 2015MNRAS.454.2855V, 2017MNRAS.470...48V}, but we have discussed that their contribution is low compared to the viscous disc. There is no UV/optical emission after the source went into quiescence, so it is unlikely that the UV/optical emission comes from a companion star.

The values of the power-law index in the above three models are obtained, assuming that the UV/optical emission is only from a single mechanism. In practice, the correlation obtained from the observations could not be well interpreted with a single theoretical model. For instance, in a study of 33 LMXBs, \citet{2006MNRAS.371.1334R} found the power-law index $\beta \sim0.6$ for BH-LMXBs in the hard accretion state, and they suggested that the UV/optical emission could be produced by more than one mechanism or that the exact contributions can not be determined only through the power-law index. In the study of NS-LMXB Aql X-1, \citet{2020MNRAS.493..940L} found that the $\beta$ values in the outburst decay are higher than the power-law index expected from the above three theoretical models, and they also suggested that multiple mechanisms are contributing to the UV/optical emissions observed in Aql X-1. 

\section{Conclusion}
\label{sec:conclusion}

We have presented simultaneous UV/optical and X-ray observations of the short orbital period BH-LMXB \source\ during its 12-yr long outburst, and obtain the following conclusions:

(i) Based on the UV/optical$-$X-ray flux correlations and the comparisons with theoretical models, we find that the UV/optical emission is dominated by the viscously heated accretion disc at large radii during the long-term outburst.

(ii) We find that the low-intensity peaks that appeared in the X-ray band during the outburst are likely ascribed to increased soft X-ray flux from the inner part of the accretion disc.

\section*{Acknowledgments}
{This work made use of data supplied by the UK Swift Science Data Centre at the University of Leicester. We made use of PyAstronomy (PyA)\footnote{https://github.com/sczesla/PyAstronomy}\citep{pya} in my work. This research made use of Astropy,\footnote{http://www.astropy.org} a community-developed core Python package for Astronomy \citep{astropy:2013, astropy:2018}, and Matplotlib \citep{Hunter:2007}. GB acknowledges funding support from the National Natural Science Foundation of China (NSFC) under grant Nos. U1838116 and Y7CZ181002.}

\section*{DATA AVAILABILITY}
The data underlying this article are available in the UK Swift Science Data Centre at the University of Leicester.




\bibliographystyle{mnras}
\bibliography{references}

%

\appendix

\section{OTHER PLOTS}
\begin{enumerate}
\item Figure A1: Evolution of the UV/optical flux in UVOT filters (v, b, uvw1, uvm2 and uvw2 filter) as a function of the 0.3$-$10 keV X-ray flux for the long-term outburst, similar to the panel (b) of Figure \ref{fig:u_x_HID_correlation_lc}. These solid lines are the power-law fit to the data with HR>0.7.

\item Figure A2: Evolution of the UV/optical flux in UVOT filters (v, b, u, uvw1, uvm2 and uvw2 filter) as a function of the 2$-$10 keV X-ray flux for the long-term outburst.

\item Figure A3: We plot the correlation between UV/optical and the two X-ray bands in the same frame to compare the difference of both bands. The correlations in 2$-$10 keV are slightly steeper than that in 0.3$-$10 keV, and the power-law indices from both bands are consistent with that predicted by the viscously heated disc.

\item Figure A4: We plot the figures similar to Figure \ref{fig:changes_in_u_x_correlation}. The purple solid line is a power-law fit to the purple points.
\end{enumerate}

\begin{figure*}
\begin{center}
\subfigure{\includegraphics[width=0.94\columnwidth]{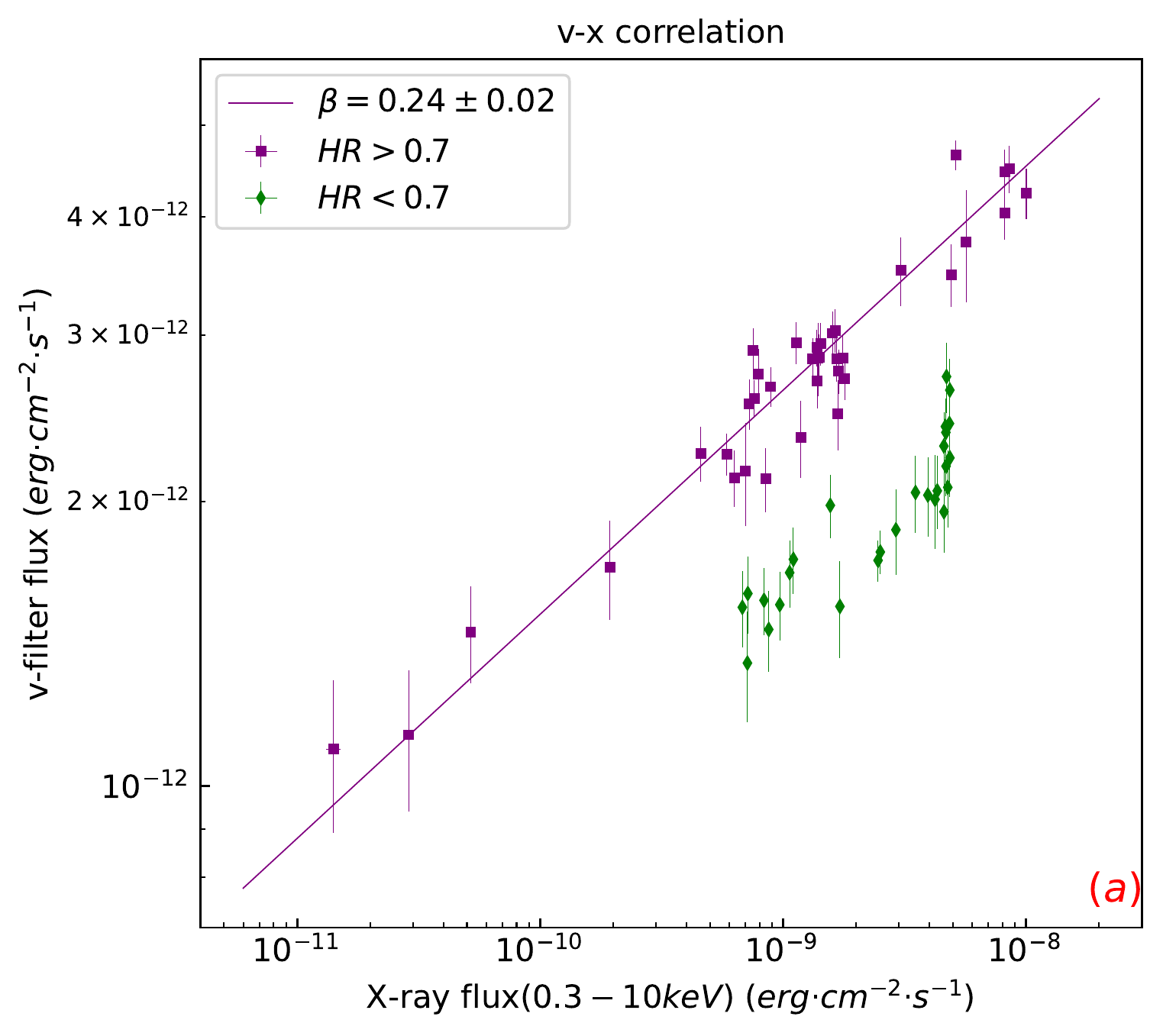}}
\subfigure{\includegraphics[width=0.94\columnwidth]{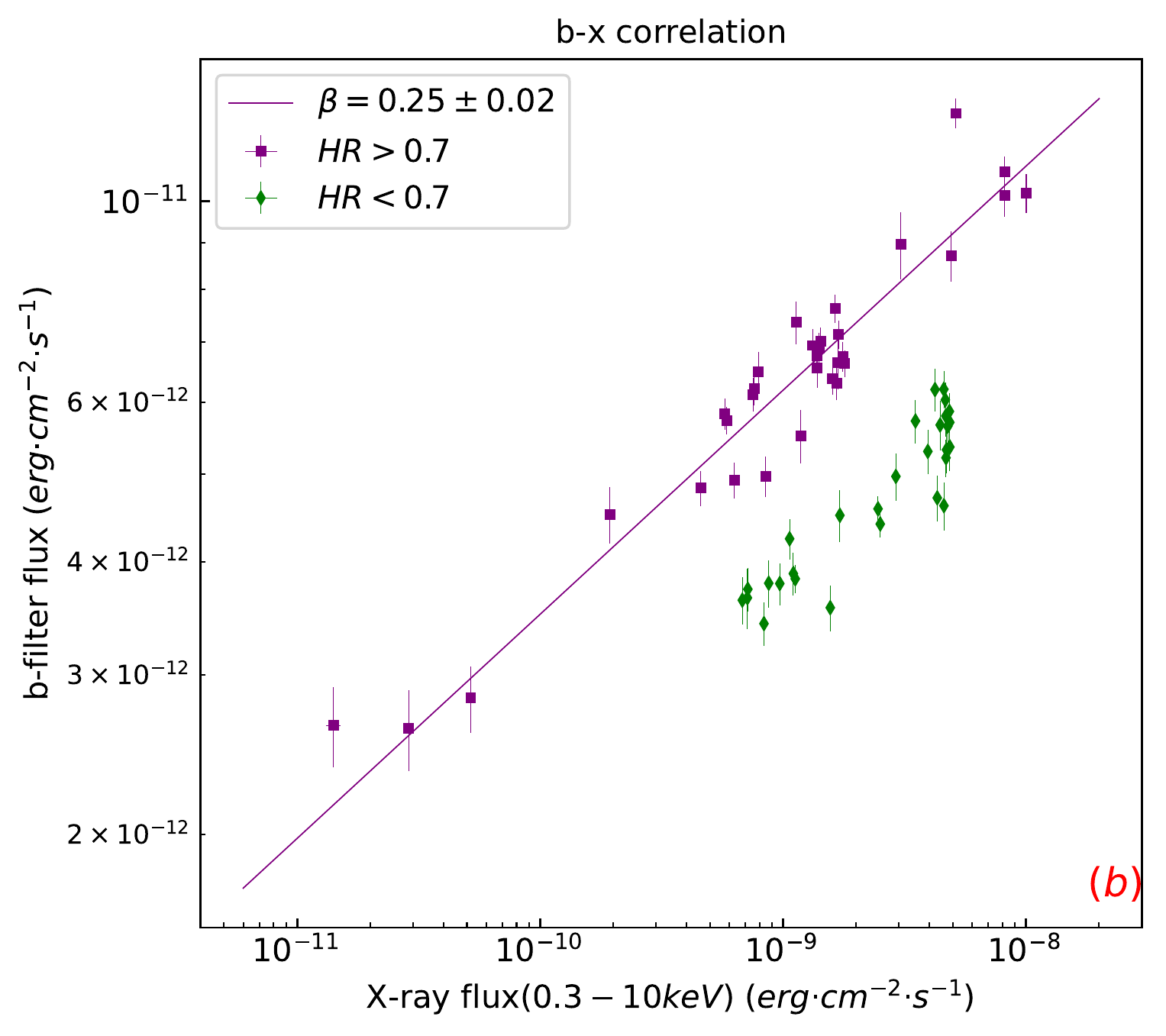}}
\subfigure{\includegraphics[width=0.94\columnwidth]{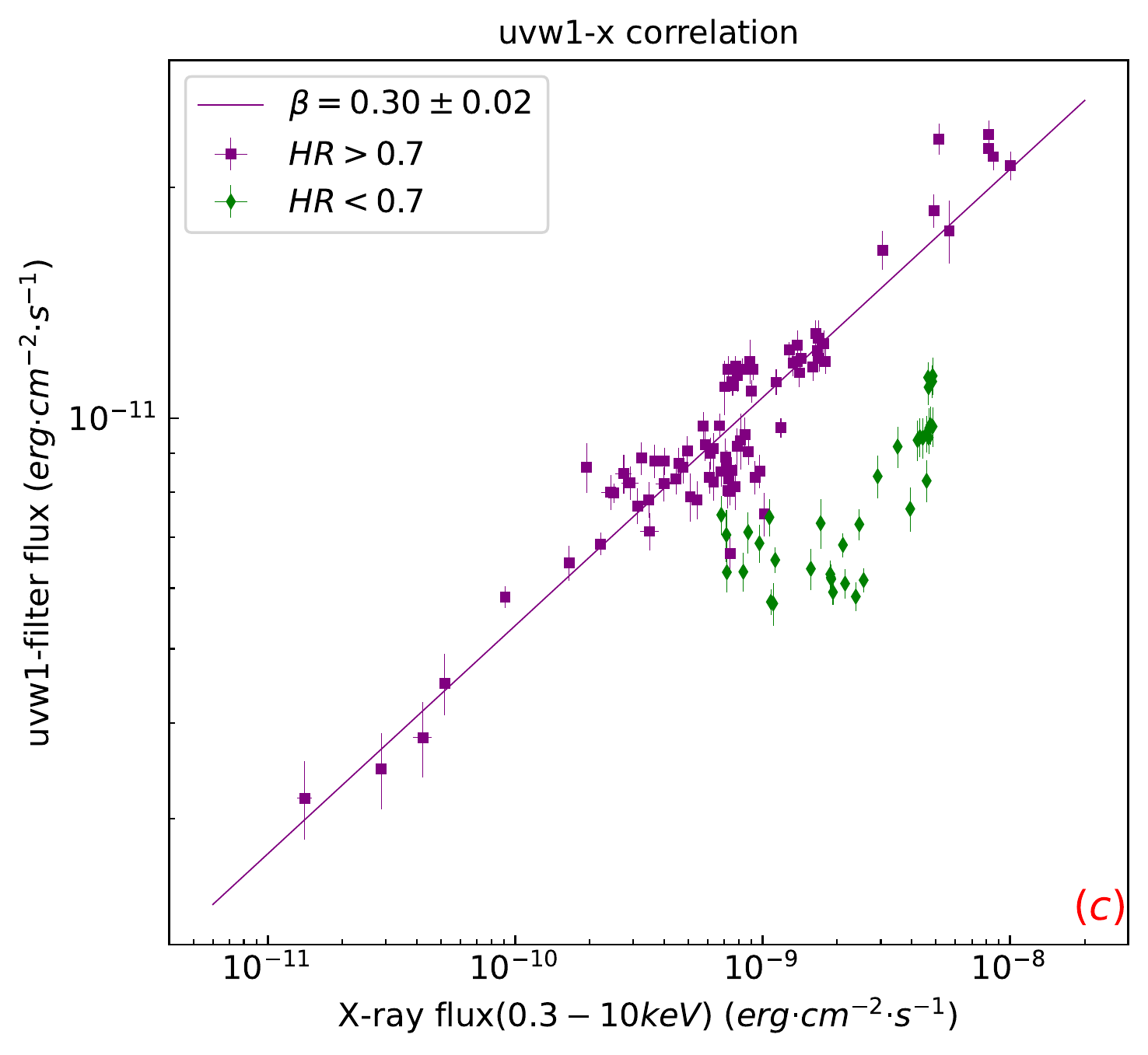}}
\subfigure{\includegraphics[width=0.94\columnwidth]{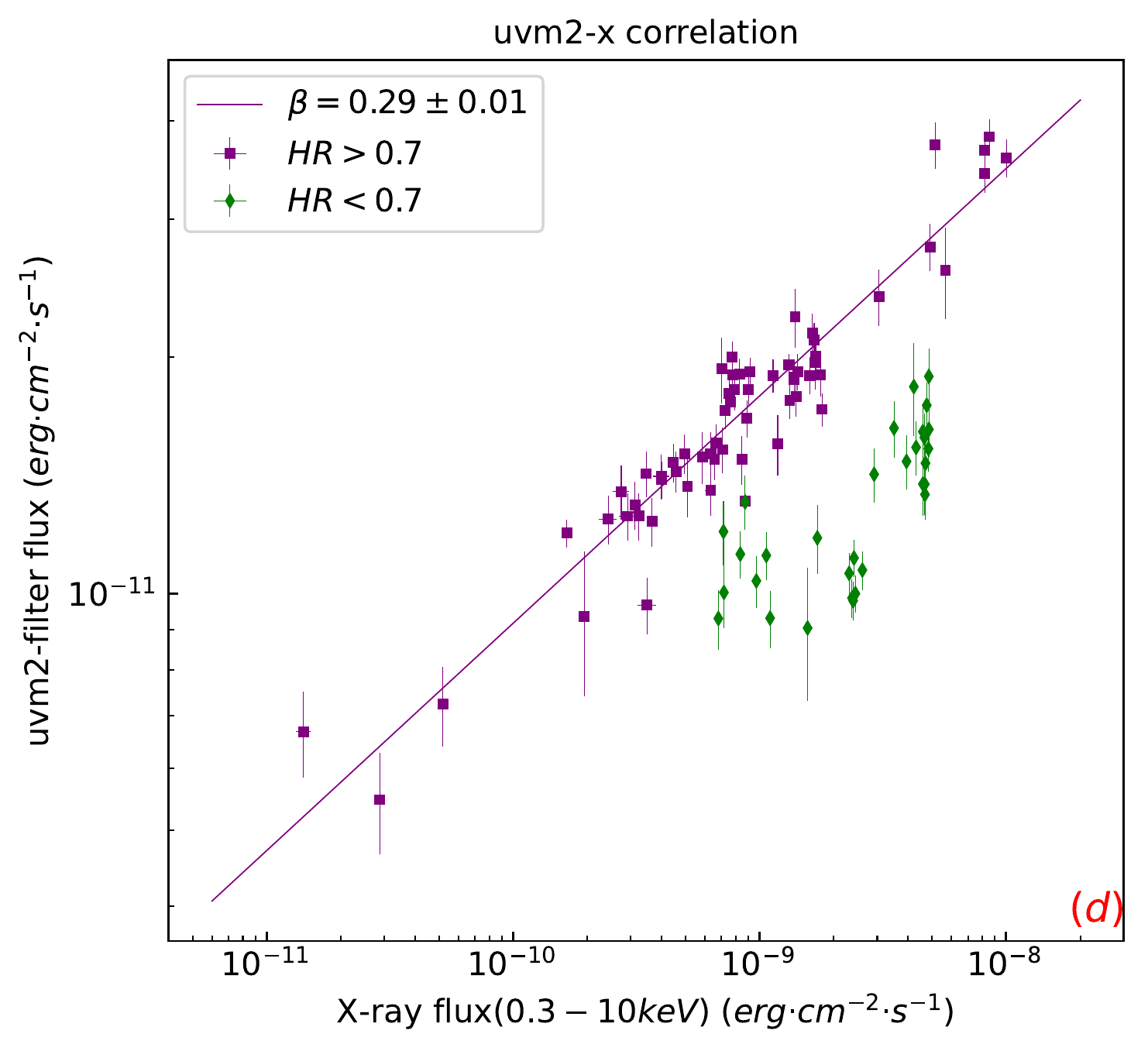}}
\subfigure{\includegraphics[width=0.94\columnwidth]{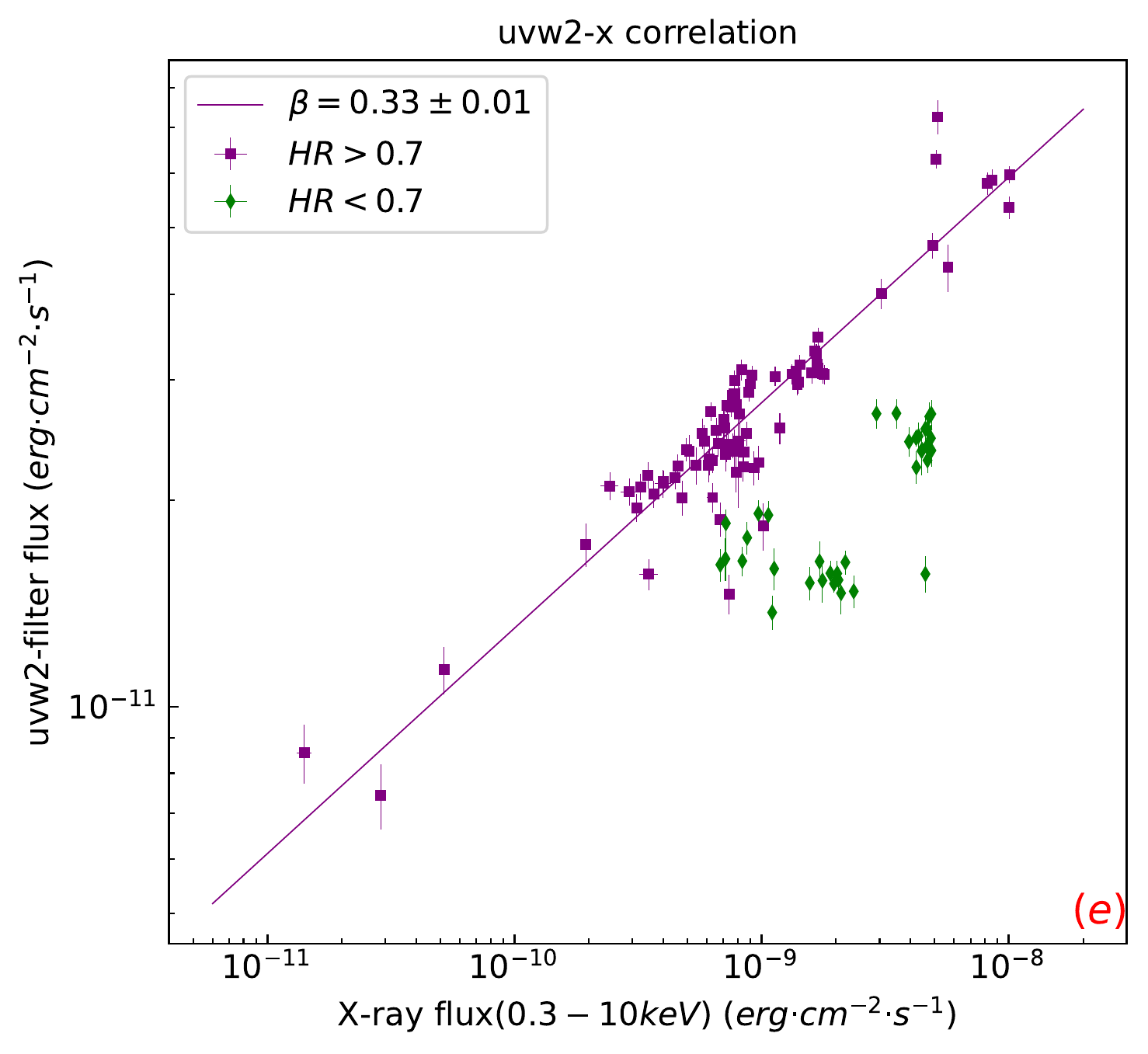}}
\caption{Other UVOT filter fluxes versus 0.3$-$10 keV X-ray fluxes.}
\label{fig:other filter vs 0.3-10 X}
\end{center}
\end{figure*}

\begin{figure*}
\begin{center}
\subfigure{\includegraphics[width=0.93\columnwidth]{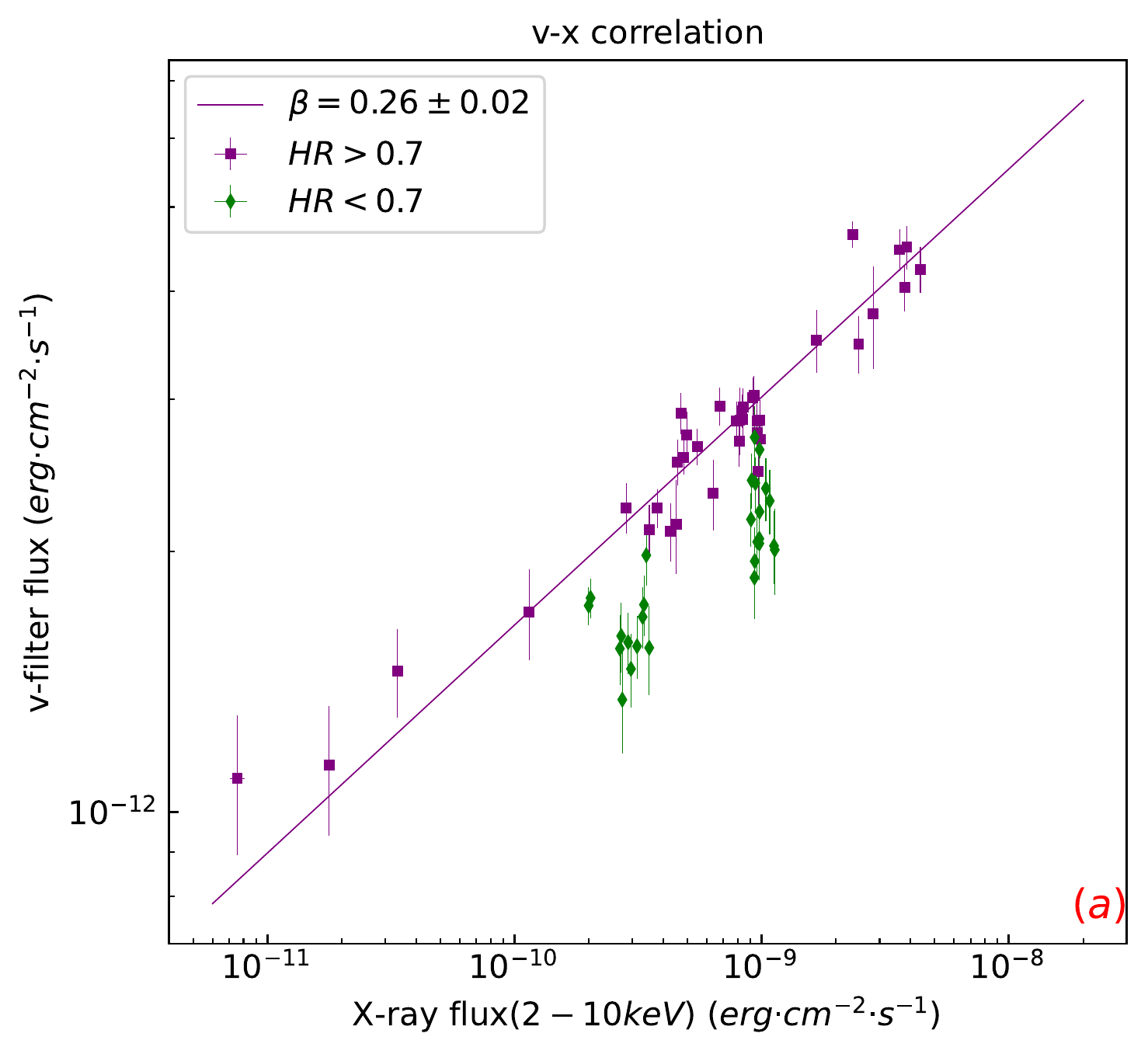}}
\subfigure{\includegraphics[width=0.93\columnwidth]{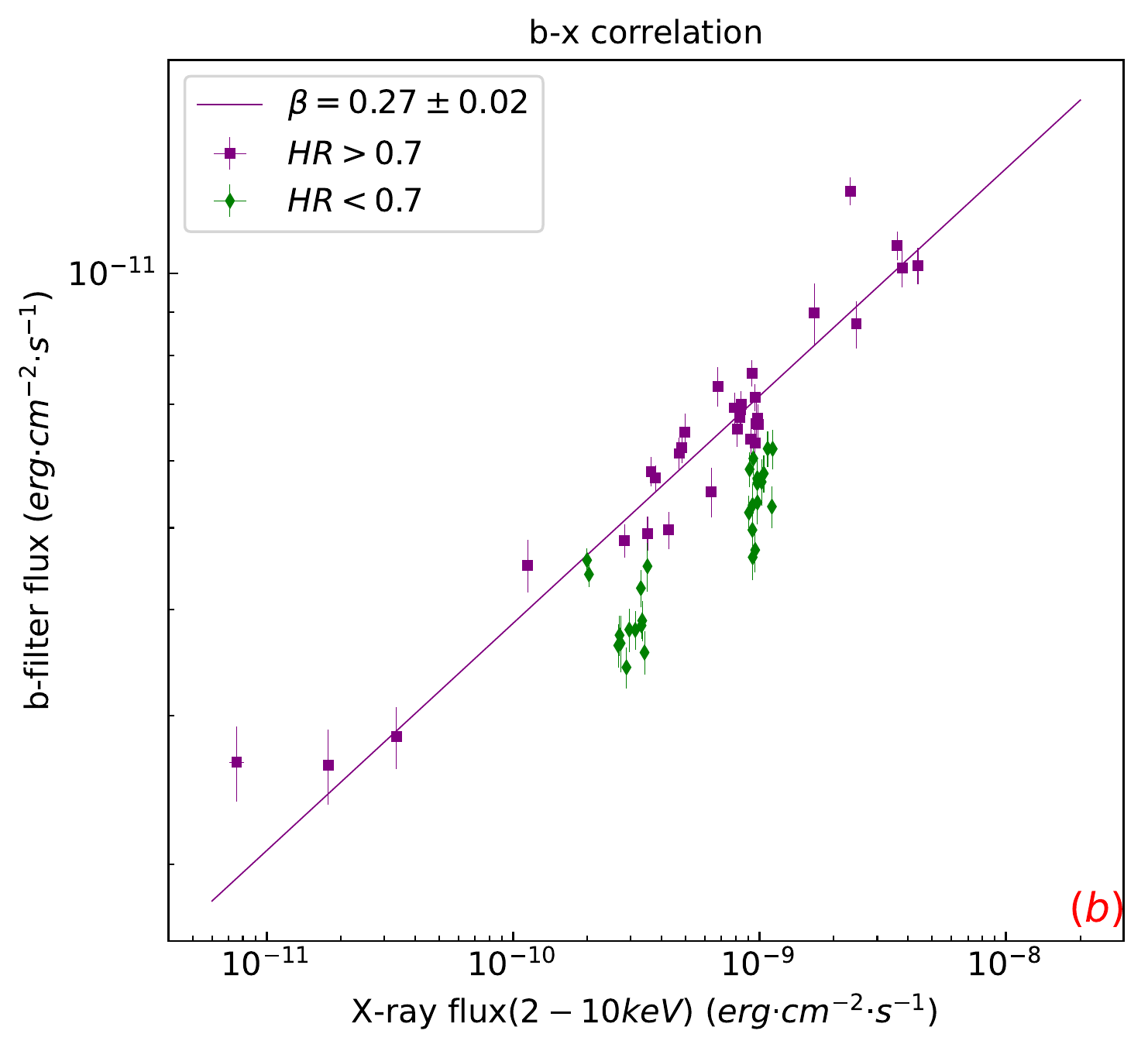}}
\subfigure{\includegraphics[width=0.93\columnwidth]{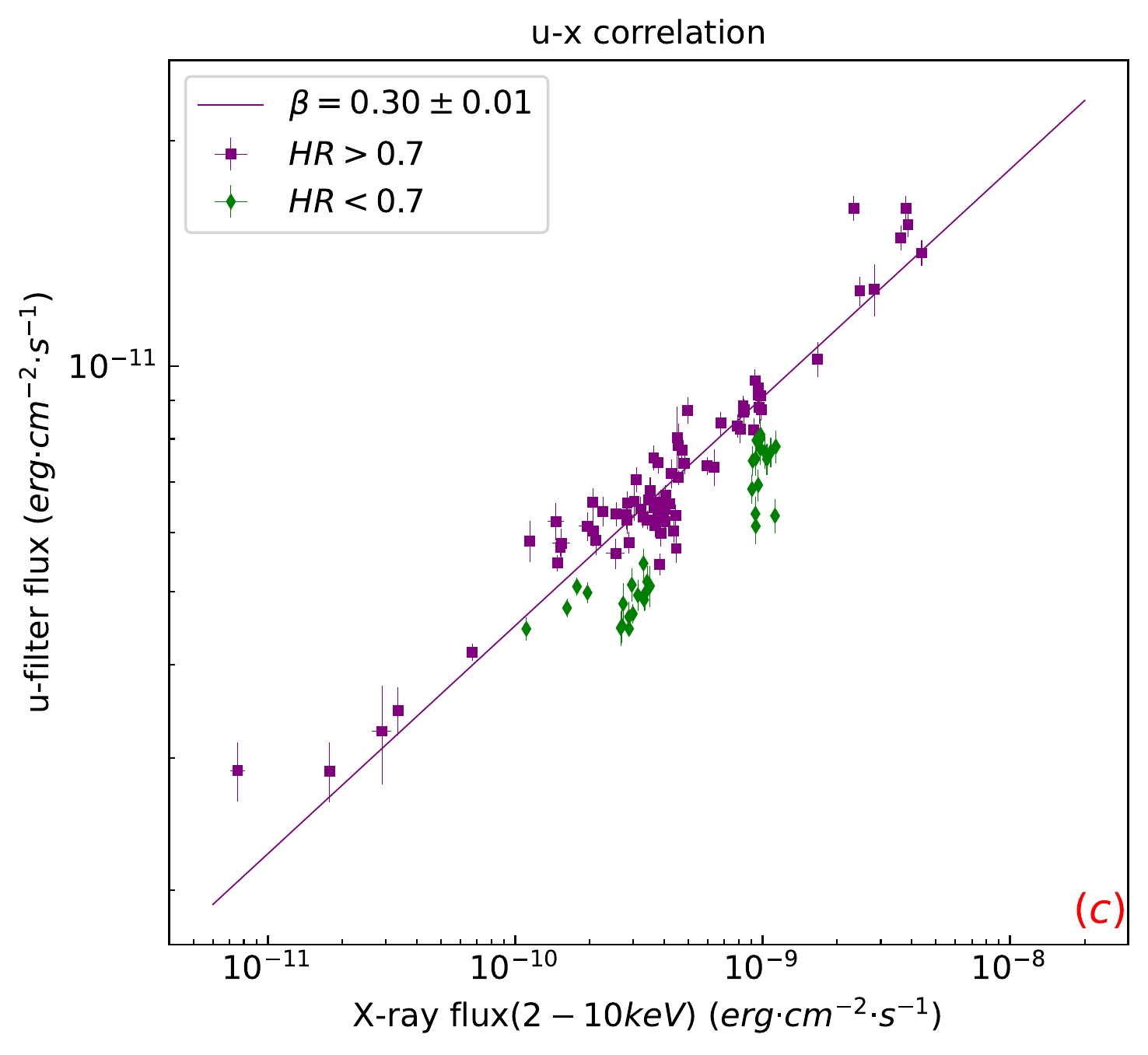}}
\subfigure{\includegraphics[width=0.93\columnwidth]{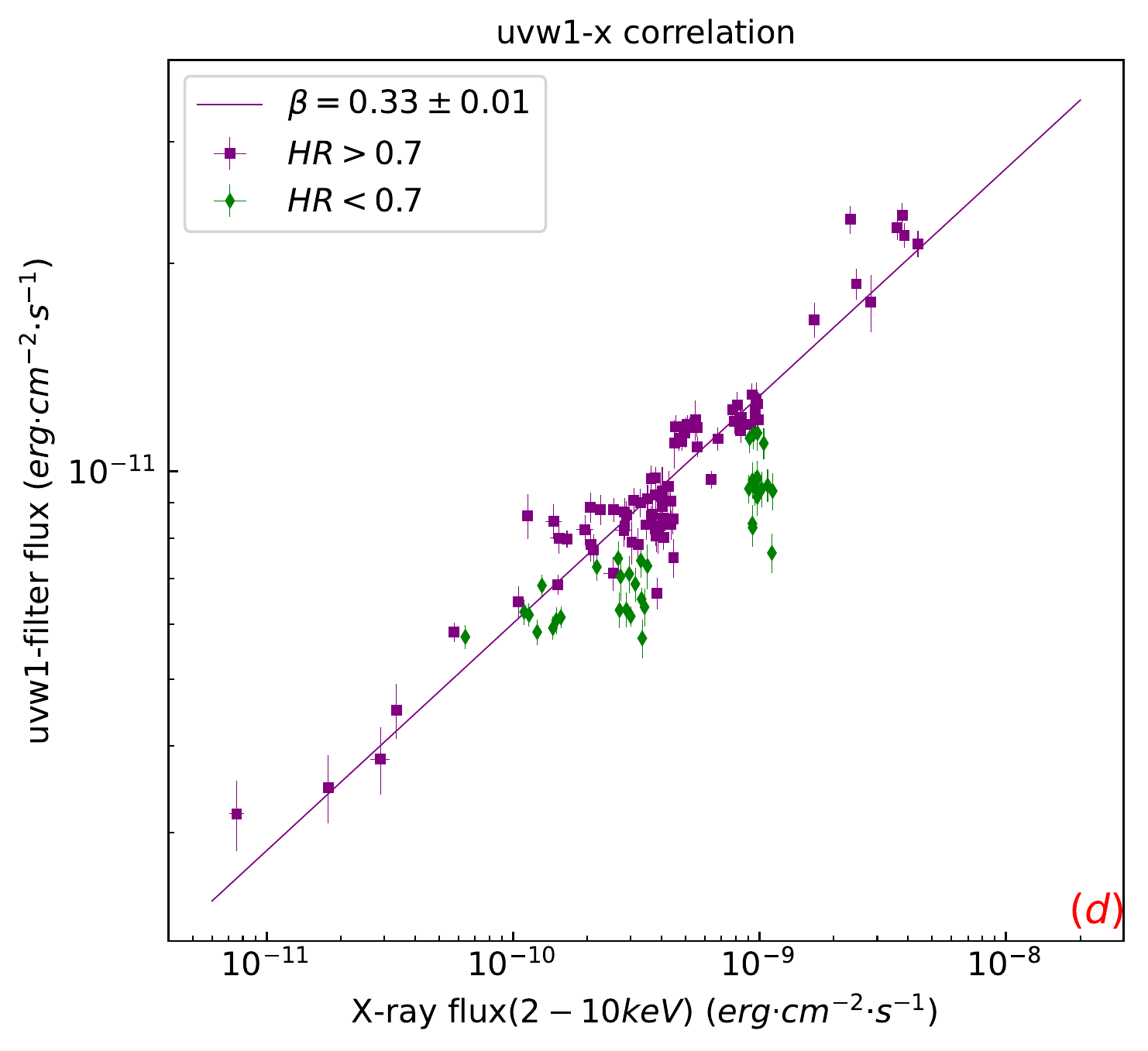}}
\subfigure{\includegraphics[width=0.93\columnwidth]{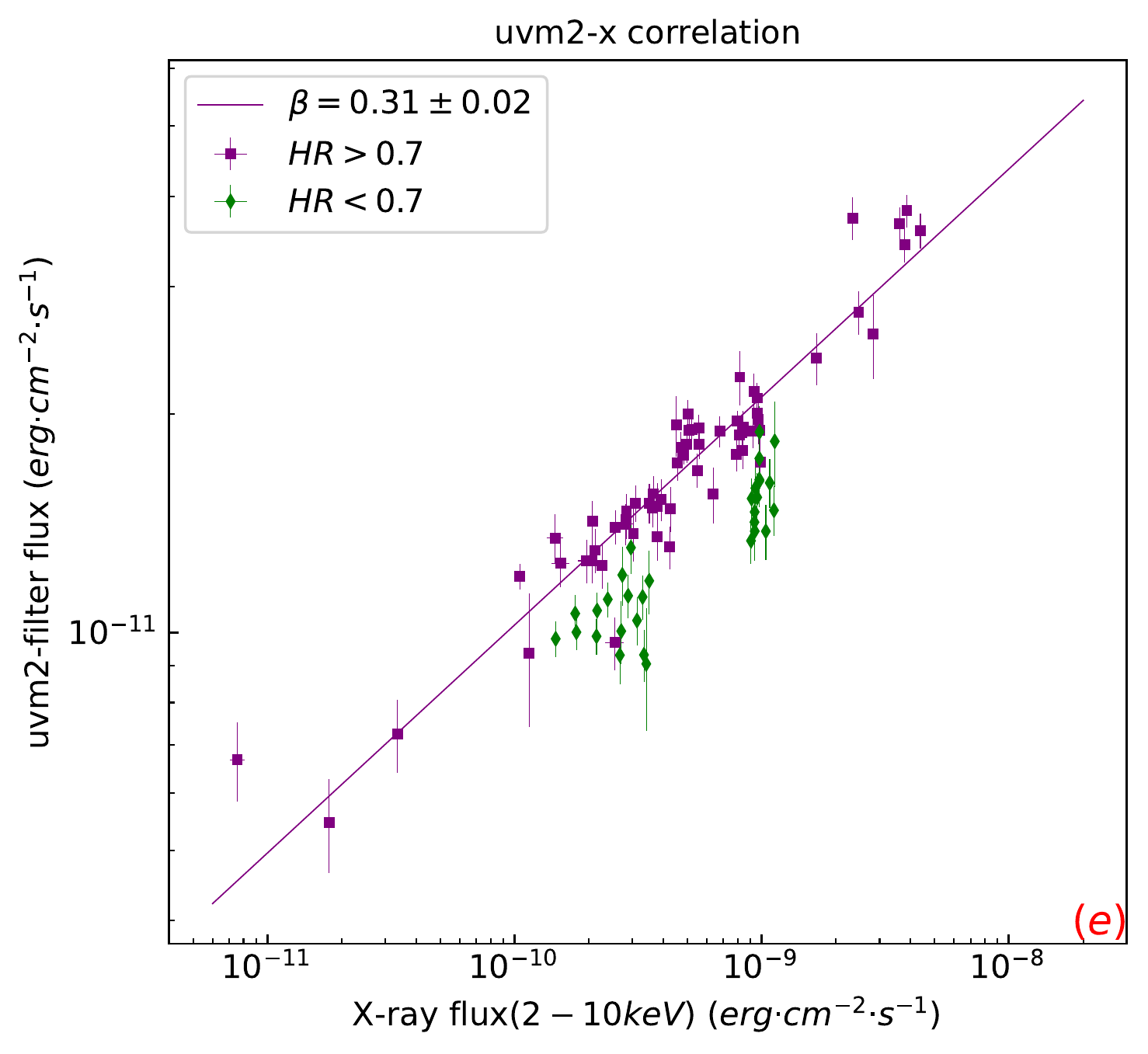}}
\subfigure{\includegraphics[width=0.93\columnwidth]{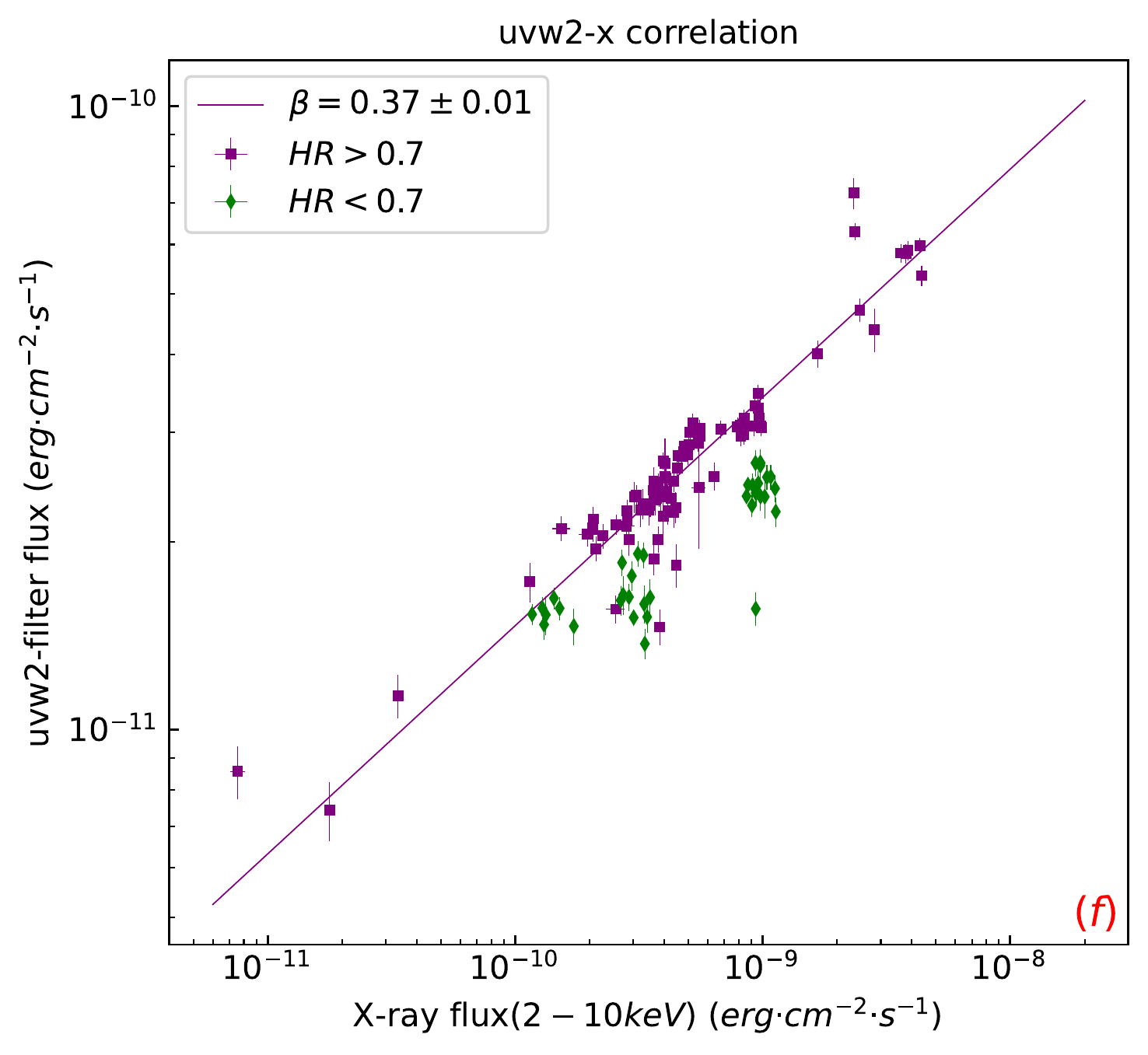}}
\caption{UVOT fluxes fluxes versus 2$-$10 keV X-ray fluxes.}
\label{fig:uvot filter vs 2-10 X}
\end{center}
\end{figure*}

\begin{figure*}
\subfigure{\includegraphics[width=0.88\columnwidth]{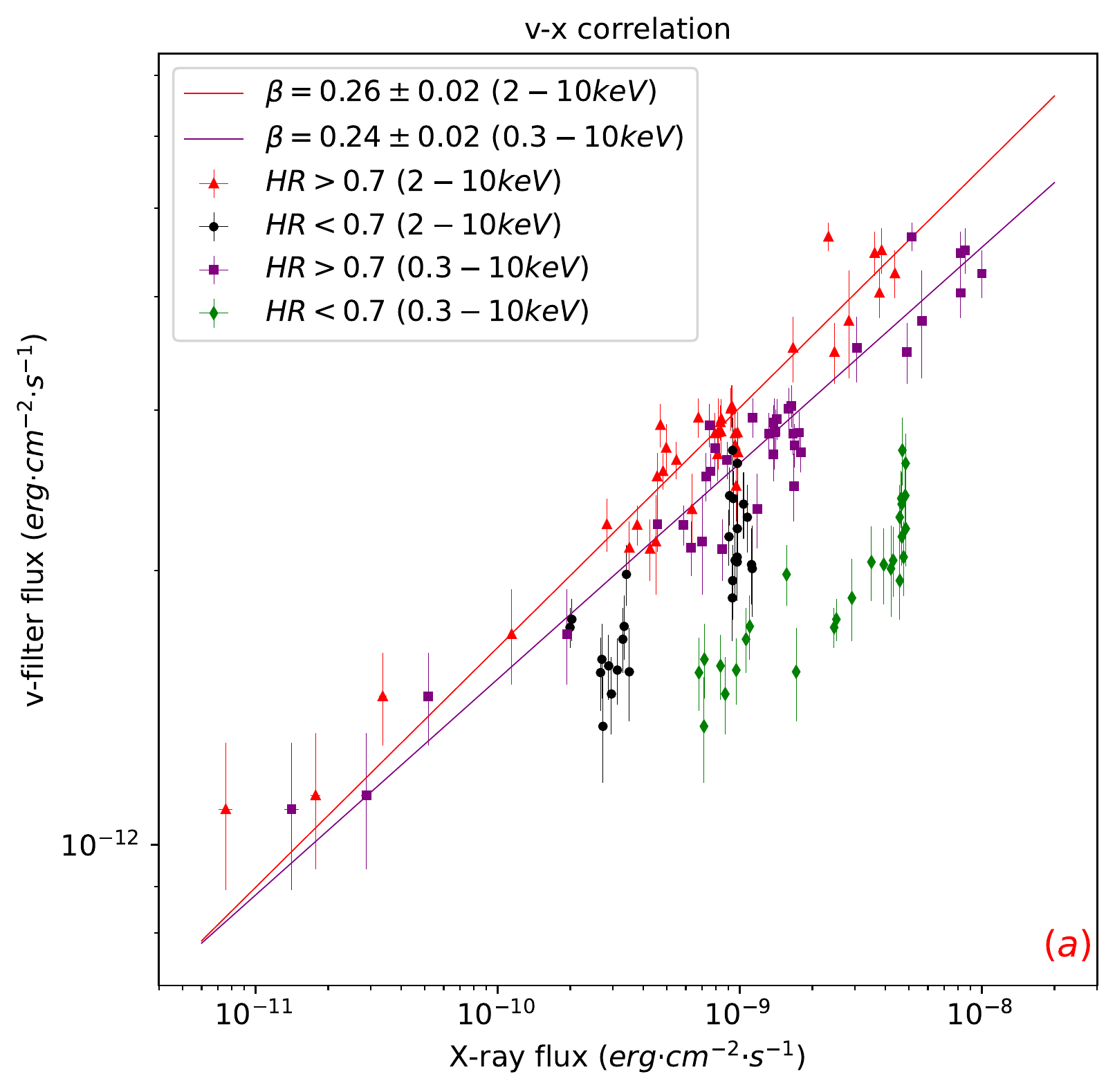}}
\subfigure{\includegraphics[width=0.88\columnwidth]{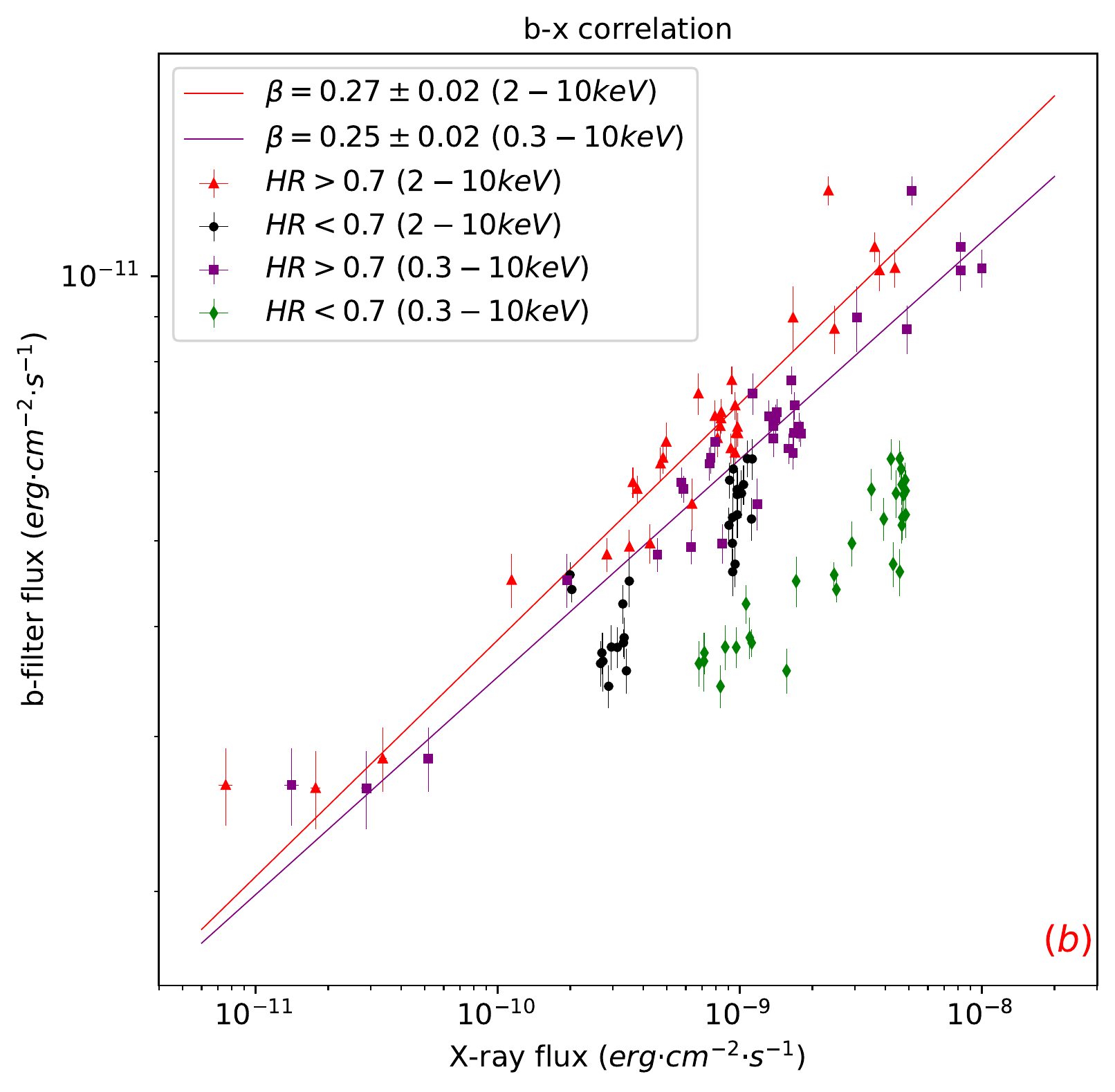}}
\subfigure{\includegraphics[width=0.88\columnwidth]{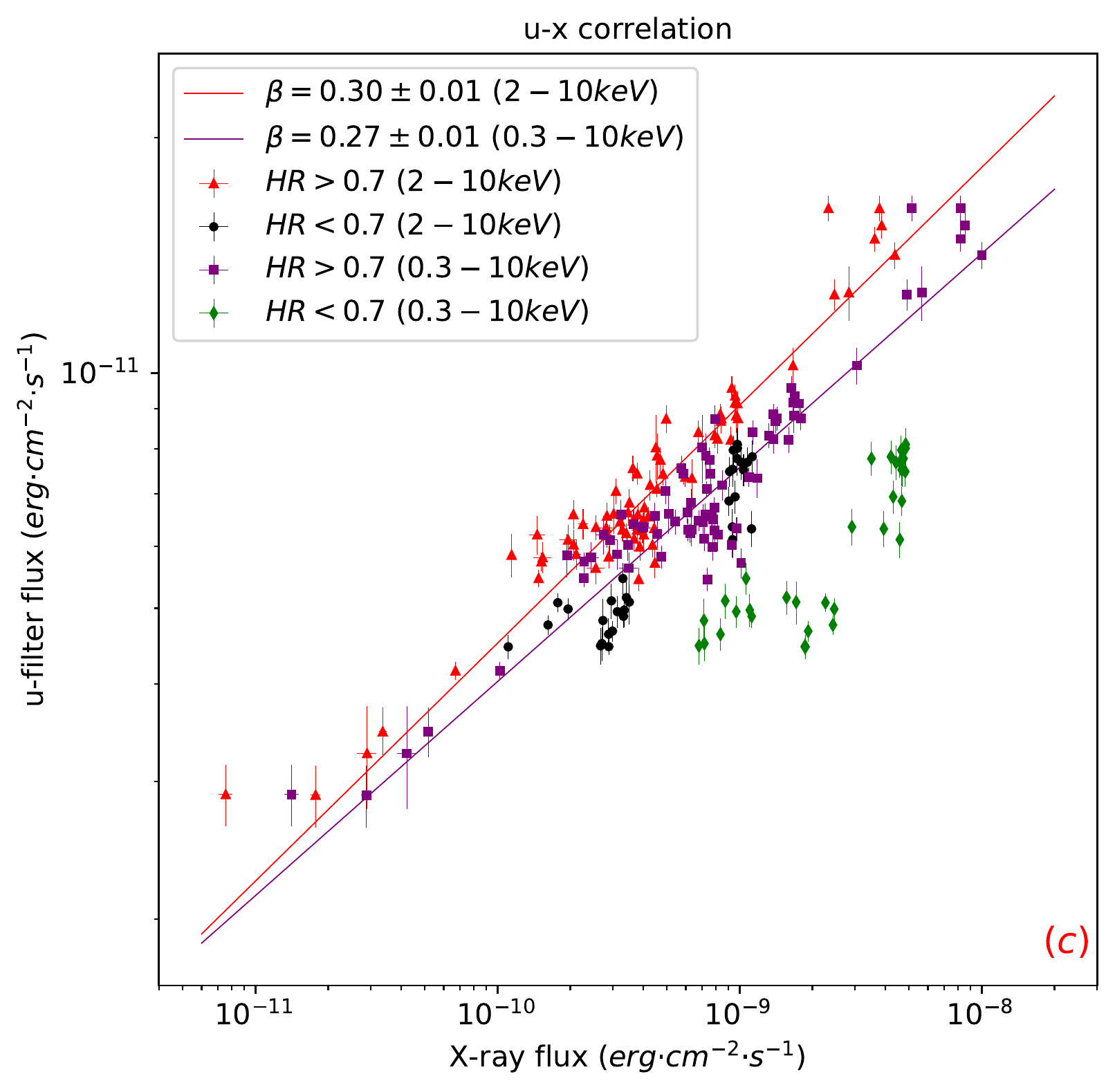}}
\subfigure{\includegraphics[width=0.88\columnwidth]{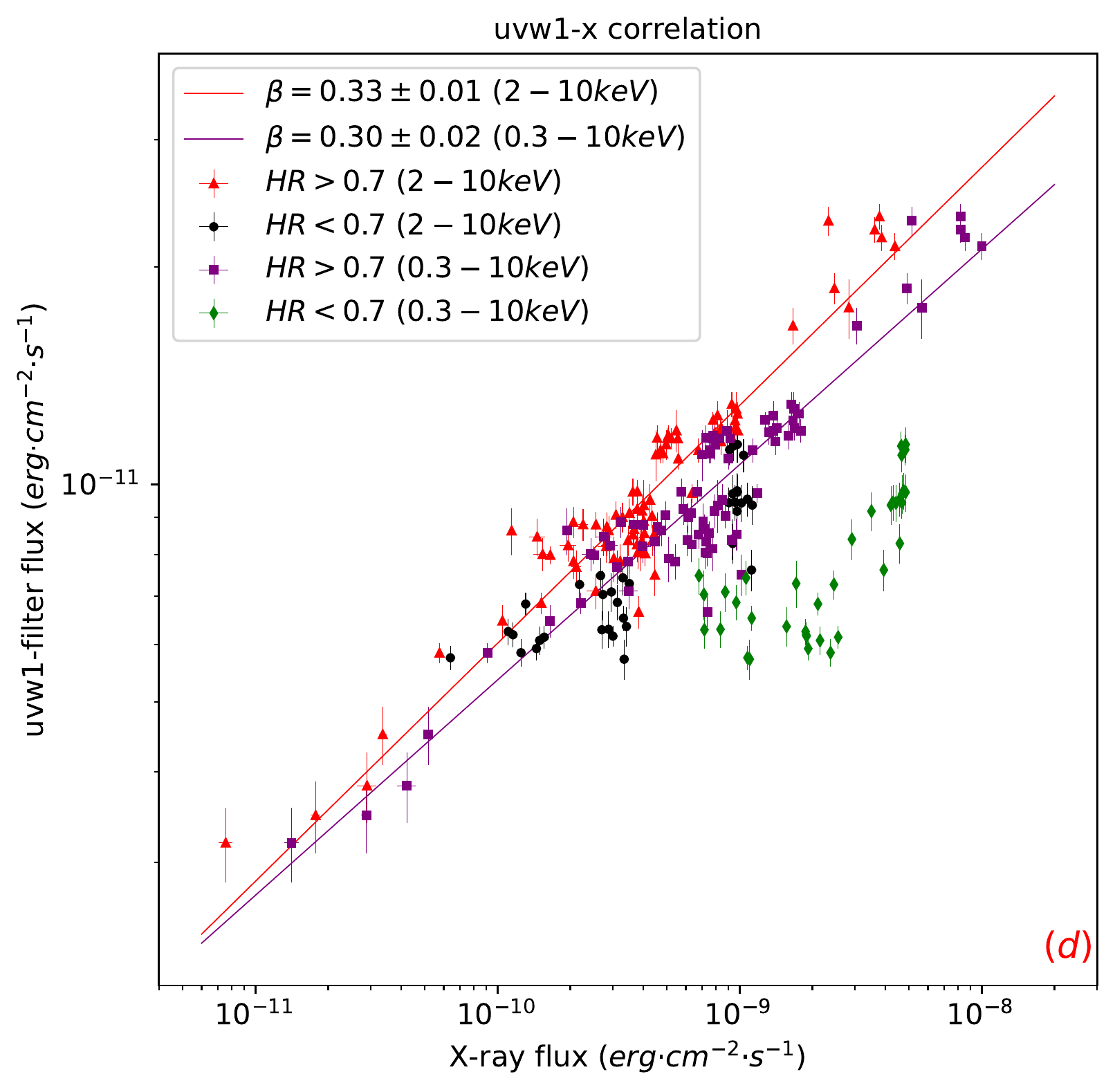}}
\subfigure{\includegraphics[width=0.88\columnwidth]{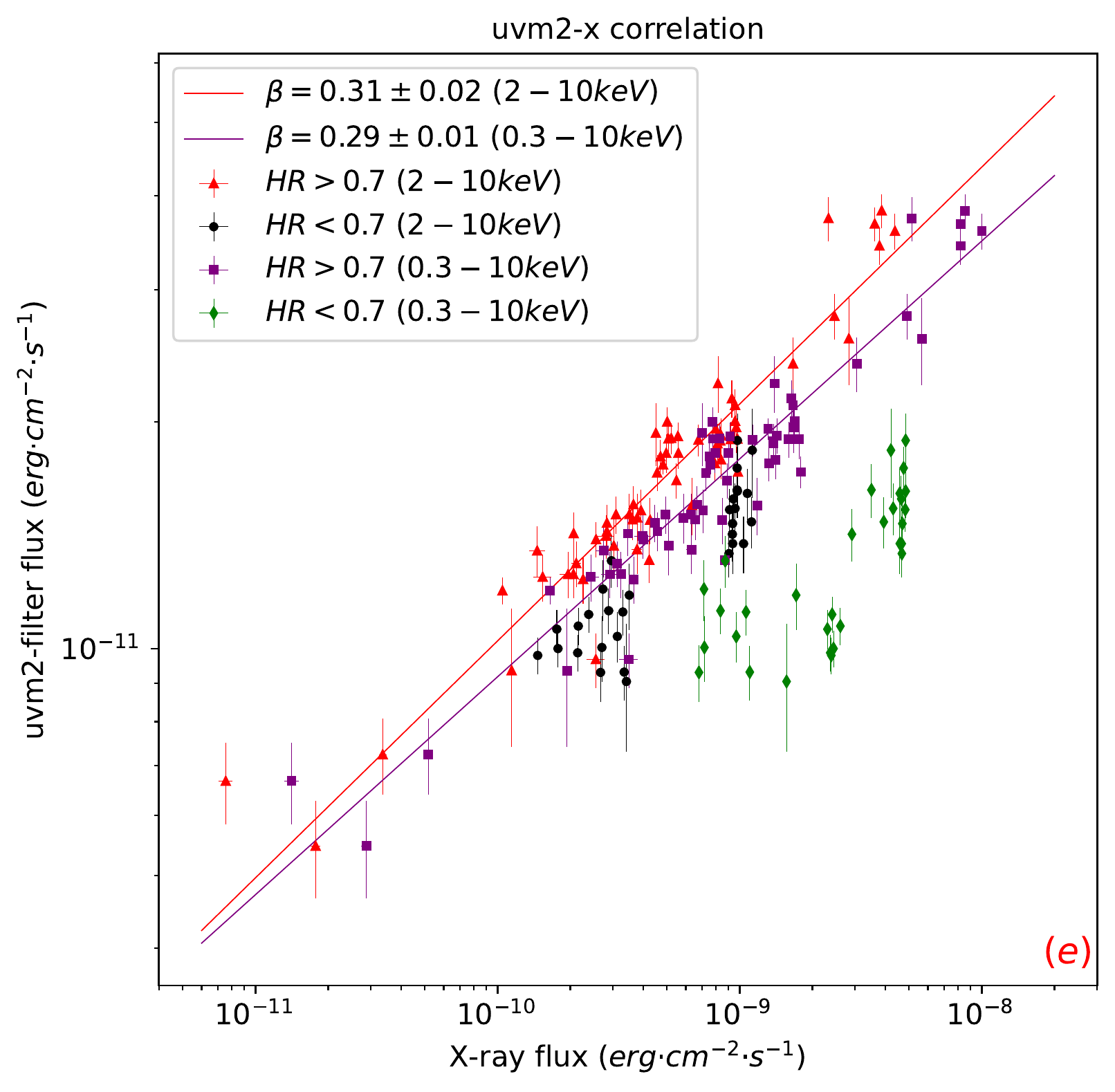}}
\subfigure{\includegraphics[width=0.88\columnwidth]{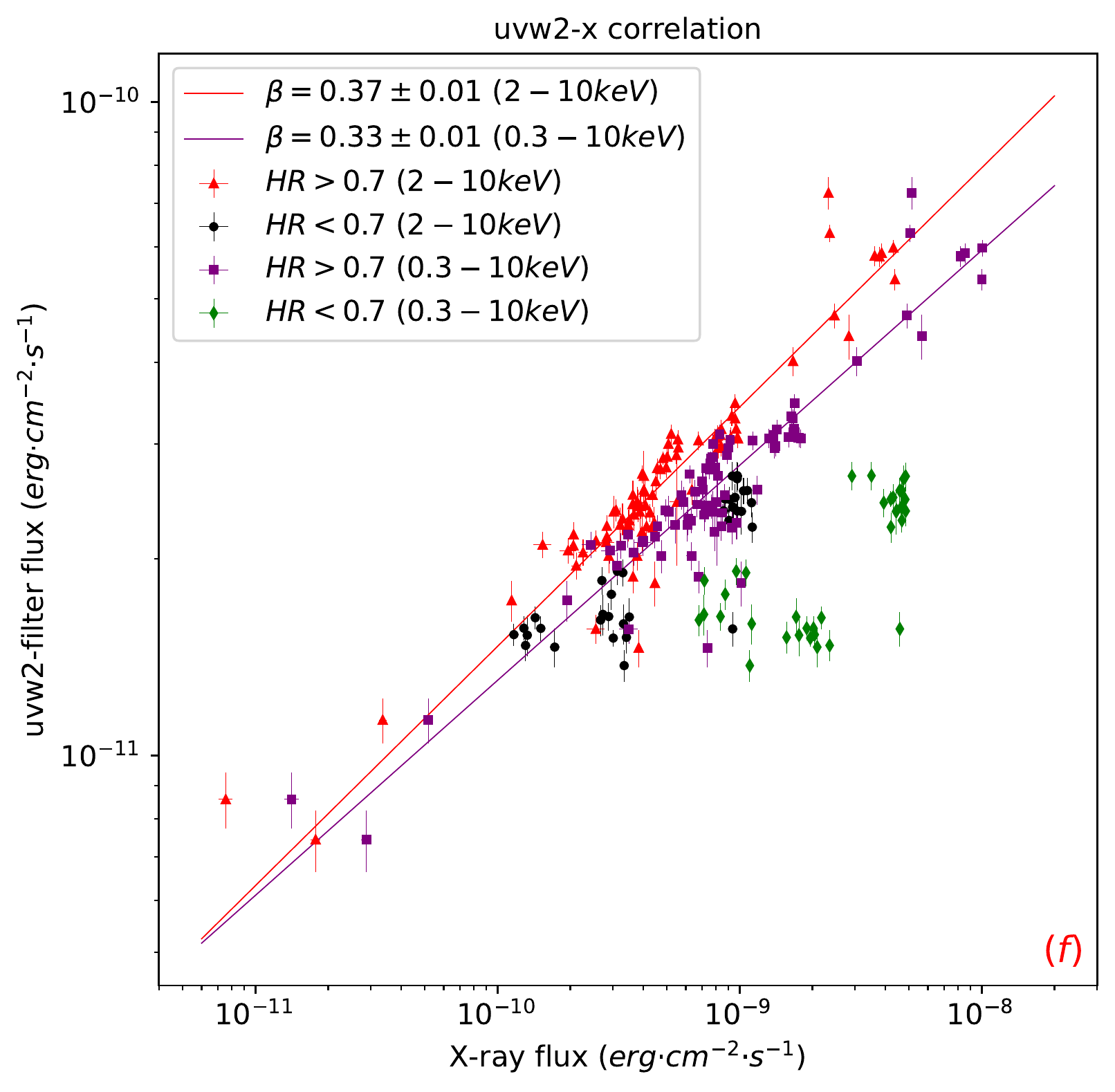}}
\caption{Comparing the correlation of different X-ray bands in the same frame.}
\label{fig:x_two_band_correlation}
\end{figure*}
\begin{figure*}
\begin{center}
\subfigure{\includegraphics[width=0.94\columnwidth]{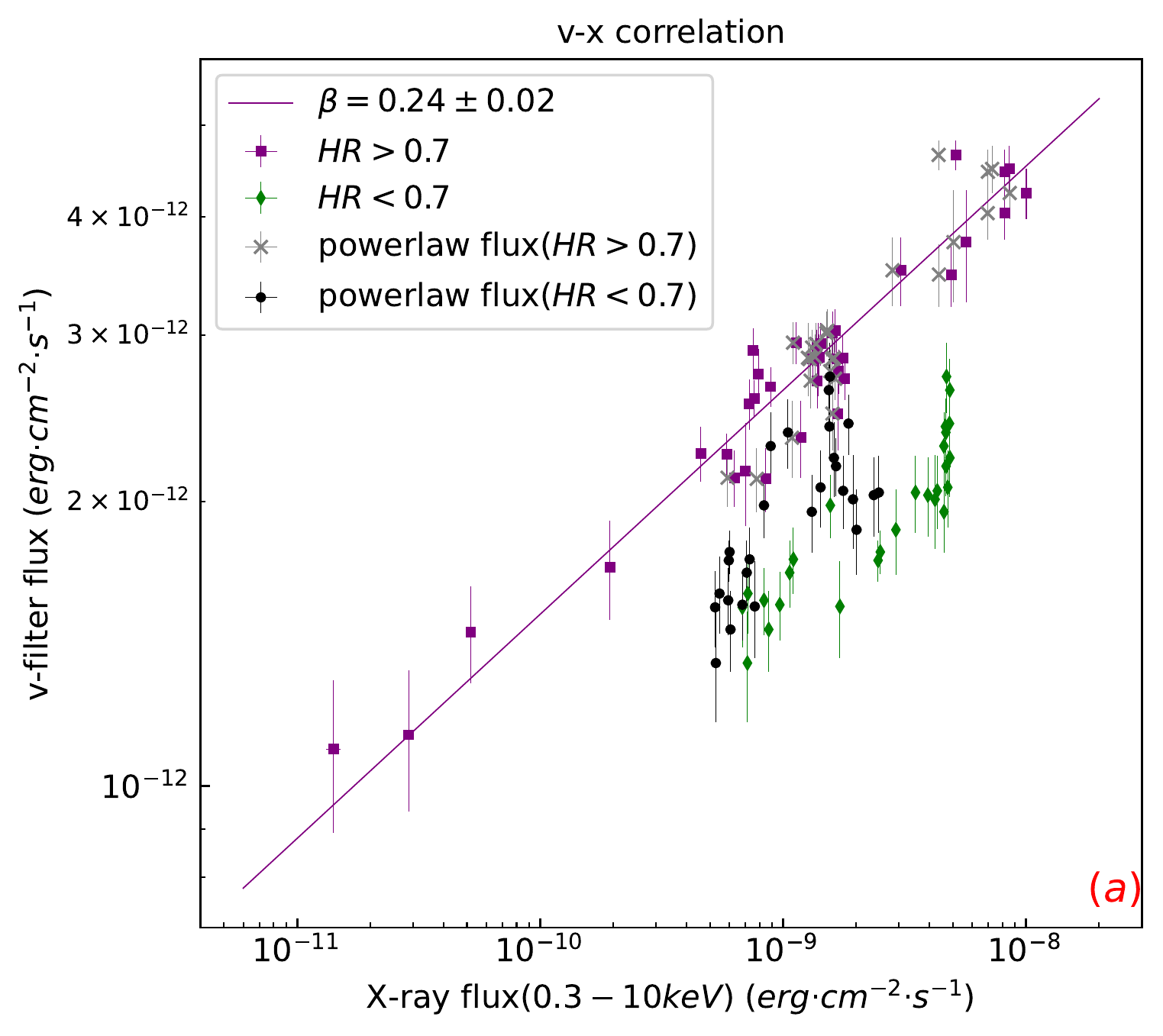}}
\subfigure{\includegraphics[width=0.94\columnwidth]{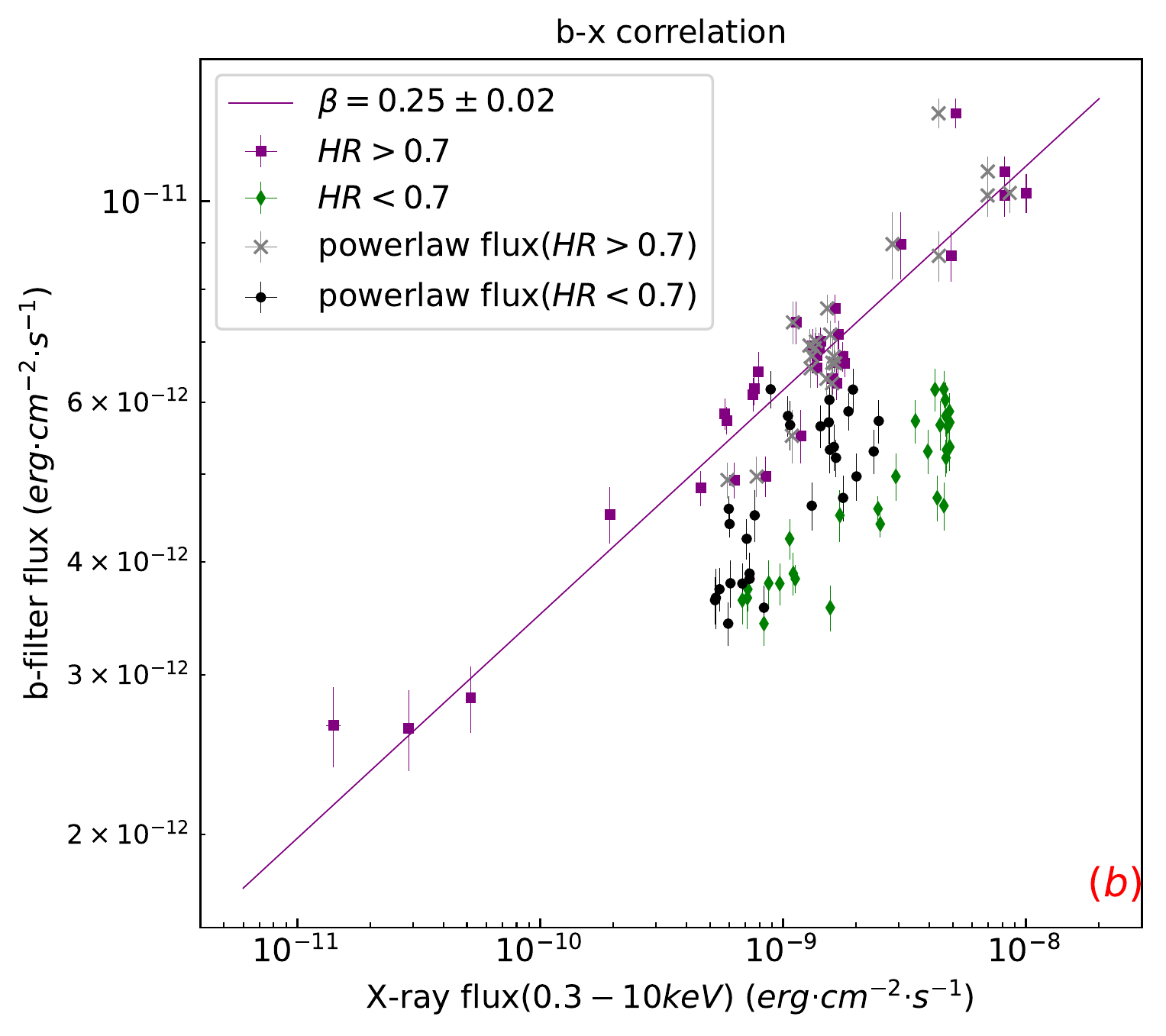}}
\subfigure{\includegraphics[width=0.94\columnwidth]{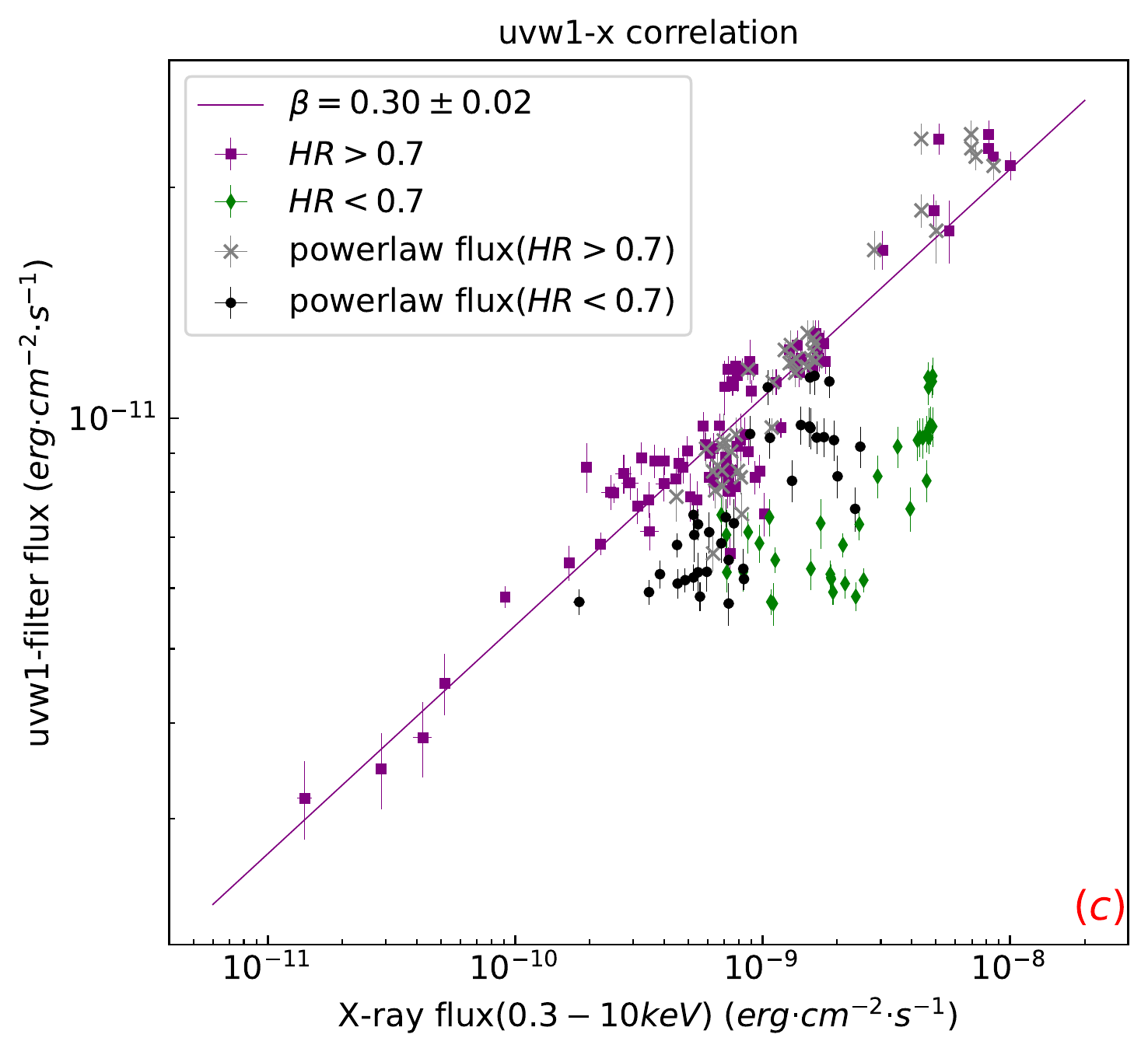}}
\subfigure{\includegraphics[width=0.94\columnwidth]{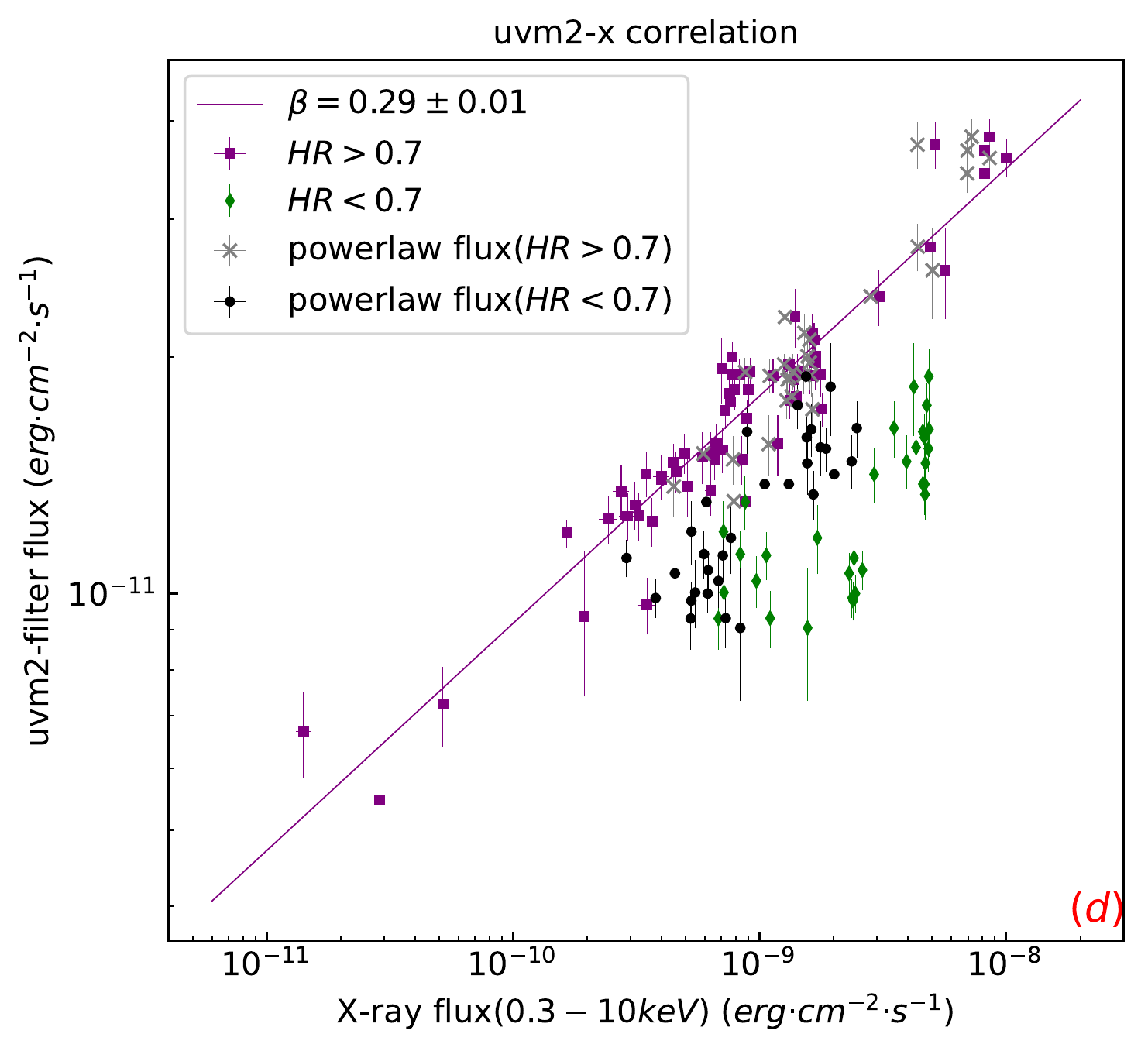}}
\subfigure{\includegraphics[width=0.94\columnwidth]{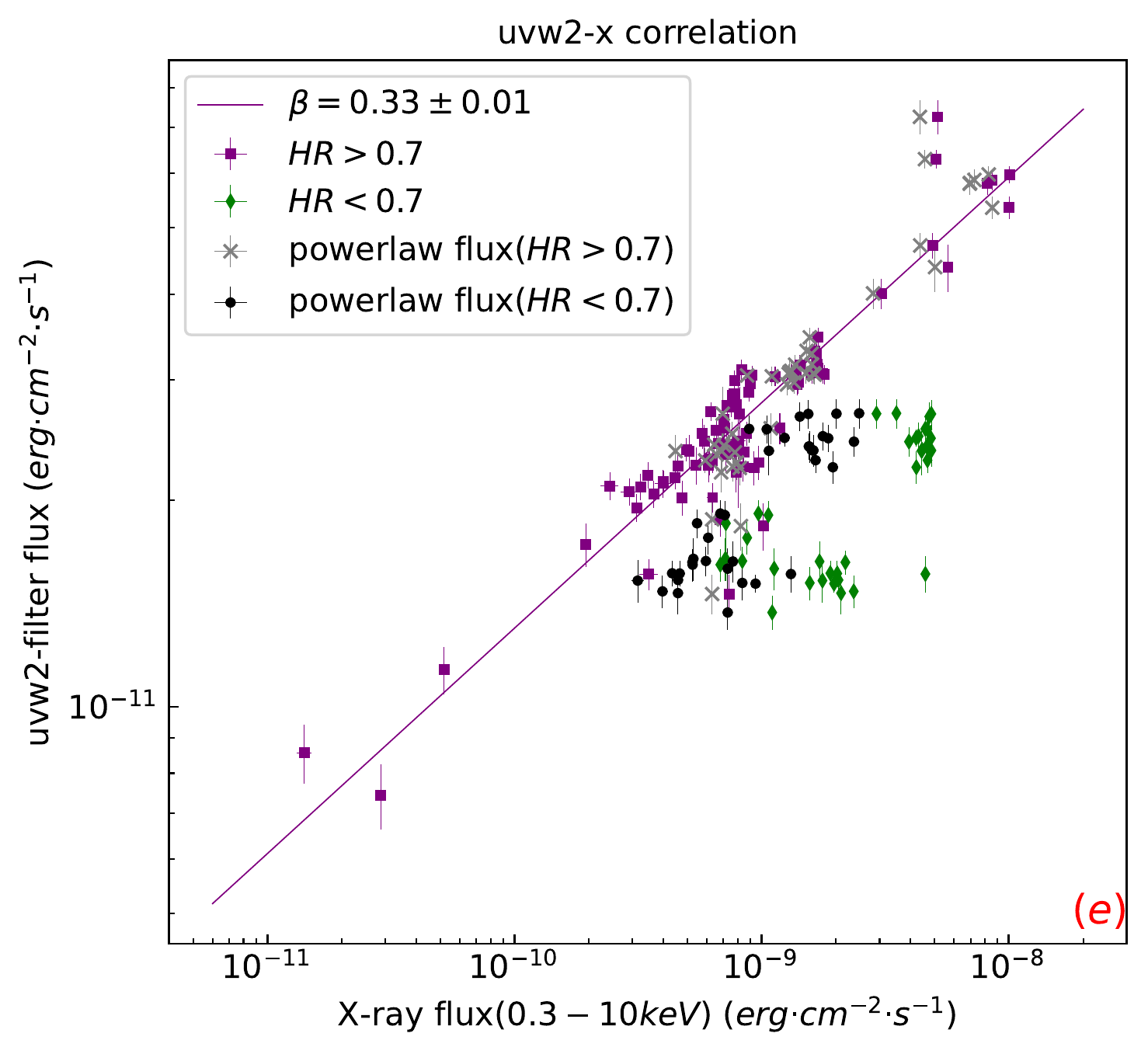}}
\caption{The flux correlation subtracted the flux of the \texttt{diskbb} component.}
\label{fig:changes_of_corr_after_del_disc_flux}
\end{center}
\end{figure*}

\bsp	
\label{lastpage}

\end{document}